\documentclass[journal,10pt]{IEEEtran}

\usepackage{graphicx, epstopdf}
\usepackage{amssymb}
\usepackage{amsmath}
\usepackage{amsthm}
\usepackage{mathrsfs}
\usepackage{citesort}
\usepackage{color}
\usepackage[T1]{fontenc}
\usepackage[latin9]{inputenc}
\usepackage{array}
\usepackage{textcomp}
\usepackage{multirow}
\usepackage{soul}
\usepackage[short]{optidef} 
\usepackage{cleveref} 
\usepackage[linesnumbered, ruled]{algorithm2e}

\newtheorem{theorem}{Theorem}

\newtheorem{remark}{Remark}
\newtheorem{proposition}{Proposition}

\ifCLASSINFOpdf
  
\else
 
\fi

\begin{document}
\bstctlcite{IEEEexample:BSTcontrol}
%
\title{Short-Packet Communications for MIMO NOMA Systems over Nakagami-\textit{m} Fading: BLER and Minimum Blocklength Analysis}

\author{ Duc-Dung Tran,~\IEEEmembership{Student Member,~IEEE,}
        Shree Krishna Sharma,~\IEEEmembership{Senior Member,~IEEE,}
		Symeon Chatzinotas,~\IEEEmembership{Senior Member,~IEEE,}
		Isaac Woungang,~\IEEEmembership{Senior Member,~IEEE,}
		and~Bj\"orn Ottersten,~\IEEEmembership{Fellow,~IEEE}
\thanks{This work was supported by the National Research Fund (FNR), Luxembourg under the CORE Project 5G-Sky.}
\thanks{
D.-D. Tran, S. K. Sharma, S. Chatzinotas, and B. Ottersten are with Interdisciplinary Center for Security, Reliability and Trust (SnT), University of Luxembourg, 4365 Esch-sur-Alzette, Luxembourg (e-mail: duc.tran@uni.lu; shree.sharma@uni.lu; Symeon.Chatzinotas@uni.lu; bjorn.ottersten@uni.lu).}
\thanks{I. Woungang is with the Department of Computer Science, Ryerson University, Toronto, ON M5B 2K3, Canada (email: iwoungan@ryerson.ca).}
}

\maketitle

\begin{abstract}
Recently, ultra-reliable and low-latency communications (URLLC) using short-packets has been proposed to fulfill the stringent requirements regarding reliability and latency of emerging applications in 5G and beyond networks. In addition, multiple-input multiple-output non-orthogonal multiple access (MIMO NOMA) is a potential candidate to improve the spectral efficiency, reliability, latency, and connectivity of wireless systems. In this paper, we investigate short-packet communications (SPC) in a multiuser downlink MIMO NOMA system over Nakagami-\textit{m} fading, and propose two antenna-user selection methods considering two clusters of users having different priority levels. In contrast to the widely-used long data-packet assumption, the SPC analysis requires the redesign of the communication protocols and novel performance metrics. Given this context, we analyze the SPC performance of MIMO NOMA systems using the average block error rate (BLER) and minimum blocklength, instead of the conventional metrics such as ergodic capacity and outage capacity. More specifically, to characterize the system performance regarding SPC, asymptotic (in the high signal-to-noise ratio regime) and approximate closed-form expressions of the average BLER at the users are derived. Based on the asymptotic behavior of the average BLER, an analysis of the diversity order, minimum blocklength, and optimal power allocation is carried out. The achieved results show that MIMO NOMA can serve multiple users simultaneously using a smaller blocklength compared with MIMO OMA, thus demonstrating the benefits of MIMO NOMA for SPC in minimizing the transmission latency. Furthermore, our results indicate that the proposed methods not only improve the BLER performance but also guarantee full diversity gains for the respective users.
\end{abstract}


\IEEEpeerreviewmaketitle

\section{Introduction}
\label{sec_Intro}

Ultra-reliable and low-latency communications (URLLC) has recently been considered as a promising technology for the 5th generation (5G) and beyond wireless networks to support novel applications with unprecedented requirements of reliability and latency \cite{Ji2018,Popovski2018,Sharma2020}. Furthermore, it is a potential solution for mission-critical Internet of Things (IoT) applications such as industrial automation, remote surgery, and vehicle-to-everything (V2X) communications, which require high reliability and low latency \cite{Sutton2019}. URLLC systems should be designed to meet the requirements of high reliability ($99.999\%$) and low latency ($1$ ms) \cite{Durisi2016}. To achieve such stringent requirements, a new transmission approach, i.e., short-packet communications (SPC), could be a promising solution. This is different from the traditional analytic methods designed to target Shannon's channel capacity using long data-packets, which are no longer suitable for low latency systems \cite{Durisi2016}. To characterize the performance of SPC, new performance metrics including block error rate (BLER) and overhead ratio (i.e., ratio of pilots to the information payload), have been introduced in the literature \cite{Pol2010,Yang2014,Mousaei2017}.

Besides, non-orthogonal multiple access (NOMA) has recently emerged as a promising technology to improve the spectral efficiency and user fairness for wireless networks \cite{Dai2015,Lei2019NOMA}. In contrast to the orthogonal multiple access (OMA) which utilizes orthogonal resources (e.g., time and frequency) to support multiple users, this technique can serve them at the same time/frequency/code by using the power domain and effective interference management methods, such as successive interference cancellation (SIC) \cite{Dai2015}. Therefore, NOMA can more effectively support massive connectivity and further improve the reliability and latency for wireless systems \cite{Dai2018,Cirik2019}. With its potential advantages, NOMA standardization has been recently studied in the 3rd Generation Partnership Project (3GPP) frameworks \cite{3GPP2016,3GPP042020,3GPP2018} including the 3GPP Release 16 \cite{3GPP2018}. Also, the latest trend is to employ NOMA in the uplink due to the emergence of IoT and machine-type communication systems \cite{3GPP2018,Sun2018,Sharma2020}. Thus, NOMA and its variations are expected to be employed in various 5G and beyond application scenarios \cite{Cirik2019,Makki2020,Bud2020NOMAD2D}.

In addition, the combination of NOMA and multiple-input multiple-output (MIMO) systems (so-called MIMO NOMA), which can significantly enhance the spectral efficiency and performance of NOMA systems, has also been investigated in recent years \cite{Chang2018NOMA,Yu2020}. The ergodic capacity analysis of MIMO NOMA systems has been considered in \cite{Zeng2017}, where the authors have proved the superiority of MIMO NOMA over MIMO OMA in terms of capacity. Furthermore, a transmit antenna selection (TAS) scheme has been proposed to reduce the complexity and improve the performance gain of MIMO NOMA systems \cite{Yu2017,Yu20181}. It is noteworthy that the above works on MIMO NOMA have been conducted under the assumption of long data-packet transmissions, which is no longer applicable for emerging URLLC applications with short data-packets in 5G and beyond networks \cite{Sharma2020,Sutton2019,Durisi2016}.

To overcome this challenge, in this paper, we propose to utilize SPC for MIMO NOMA systems to improve the reliability and latency as well as enhance the spectral efficiency and connectivity for wireless systems. Herein, suitable performance metrics for SPC including average BLER and minimum blocklength, are utilized instead of the conventional ones such as ergodic capacity and outage capacity.

\subsection{Related Works}
Recently, there have been a few works on SPC in NOMA systems, which is considered as a promising solution to enhance the reliability, latency, and connectivity for wireless networks \cite{Yu2018,Zheng2019,Lai2019,Wang2020,Xiao2019,Huang2019}. In particular, in \cite{Yu2018}, a two-user NOMA system with short-packets over Rayleigh fading channels was considered, in which the average BLER at users is derived to evaluate the system performance. In \cite{Zheng2019}, the BLER performance of a NOMA system was addressed, where stochastic geometry and Nakagami-\textit{m} fading channels are considered. In \cite{Lai2019}, X. Lai \textit{et al.} analyzed the performance of a cooperative NOMA SPC system over Rayleigh fading channels. However, the works \cite{Yu2018,Zheng2019,Lai2019} only considered single-input single-output (SISO) systems.

To exploit the benefits of multiple antennas in improving the reliability and reducing the latency for SPC in NOMA systems, the work in \cite{Wang2020} investigated a multiple-input single-output (MISO) scheme to evaluate the outage performance of a URLLC NOMA system with wireless power transfer. In \cite{Xiao2019}, MIMO NOMA for URLLC systems was considered to enhance the reliability and latency performance of the system. In this regard, a closed-form upper bound for the delay target violation probability was derived in \cite{Xiao2019} to identify the sufficient and necessary condition for the optimal transmit power. However, the analysis of average BLER and minimum blocklength was not considered in \cite{Xiao2019}. The work in both \cite{Wang2020} and \cite{Xiao2019} investigated a scenario where an $N$-antenna base station (BS) provides services to $N$ pairs of NOMA users, in which each pair of users is served by a distinct single transmit antenna. In contrast to this scenario, in \cite{Huang2019}, the combination of transmit antennas to serve a pair of users was examined in order to enhance the BLER performance of short-packet NOMA systems by utilizing the maximum ratio transmission (MRT), in which only the MISO scenario was considered.

Although MRT can significantly improve the system performance by combining all transmit antennas for transmission, it leads to high complexity of the signal processing and feedback overhead \cite{Yan2013}. Against this context, TAS has been proposed as a low-complexity and power-efficient solution for multi-antenna transmitters to enhance the performance of NOMA systems by selecting a best transmit antenna for transmission that maximizes the signal-to-noise ratio (SNR) at the receiver side \cite{Do2018,Yu20181}. Nevertheless, the short-packet transmission in MIMO NOMA systems considering the TAS solution, average BLER, and minimum blocklength has not yet been analyzed. Furthermore, it is noted that most of these existing studies \cite{Yu2018,Zheng2019,Lai2019,Wang2020,Xiao2019,Huang2019} only investigated Rayleigh fading channels. Research on SPC for MIMO NOMA systems applying TAS for the transmitter, selection combining (SC) and maximal ratio combining (MRC) for the receiver, over a generic fading channel, i.e., Nakagami-\textit{m}, to improve the system performance more effectively and bring more general insights of the system behavior has not yet been conducted, and thus is the focus of this paper.

\subsection{Contributions}
\label{sec_contributions}

In contrast to the existing related works, in this paper, we propose a new framework to analyze the system performance of utilizing SPC in a NOMA network, in which MIMO and Nakagami-\textit{m} distribution are considered. Most existing works on NOMA are conducted under the assumption that NOMA is carried out based on the difference in users' channel conditions \cite{Dai2015,Lei2019NOMA,Dai2018,Bud2020NOMAD2D,Chang2018NOMA,Yu2020,Zeng2017,Yu2017,Yu20181,Yu2018,Zheng2019,Lai2019,Wang2020,Xiao2019,Huang2019}. More precisely, in a two-user downlink NOMA system, a BS transmits information to the users by superimposing users' messages with different transmit power levels \cite{Dai2015}. The user having worse channel quality is allocated with the higher power level compared with the user having a better channel condition. However, in practice, users may have similar channel conditions but require different quality of service (QoS) as discussed in \cite{Ding2016nomaIoT,Din20162,Tran2018}. For example, some users may need to be served faster with low targeted data-rate, i.e, incident alerts, while some users can be served with the best effort, i.e., downloading of multimedia files \cite{Din20162}. In such a heterogeneous scenario, NOMA scheme becomes advantageous as compared to the conventional OMA as it can concurrently serve users having different QoS priorities with the same resources (time/frequency/code).

Given this context, we examine a scenario, in which a BS communicates with two user clusters having different priority levels over Nakagami-\textit{m} fading channels, where the BS and all users are equipped with multiple antennas. Note that Nakagami-\textit{m} is described as a general distribution that can include the well-known Rayleigh and Rician distributions. To perform NOMA, user paring is employed as discussed in \cite{Din2016,Liu2016,JWang2020} to reduce the strong co-channel interference in NOMA systems. Furthermore, different MIMO schemes are investigated to exploit the benefits of multiple antennas. Particularly, at the BS, TAS is utilized to select the best transmit antenna for transmission that maximizes the post-processed SNR at the receiver \cite{Do2018,Yu20181}. Besides, at the user-side, two different diversity techniques are investigated: 1) SC, which selects the best received signal branch for further processing; and 2) MRC, which combines all the received signal branches from receive antennas to maximize the output SNR. The main contributions of this paper are summarized as follows:

\begin{itemize}
	\item Firstly, we propose a novel framework to analyze the performance of an SPC-based NOMA system, where MIMO transmission and Nakagami-\textit{m} fading are considered. To achieve a general insight into the system behavior, we investigate two different cases of applying MIMO schemes for the transmitter and receiver sides including TAS/SC and TAS/MRC. Moreover, we investigate two antenna-user selection methods, namely high-priority cluster selection (HCS) and low-priority cluster selection (LCS), to design the effective communication protocols for SPC in a MIMO NOMA system.
	\item Secondly, we derive closed-form expressions for the average BLER of users in all considered cases. It should be noted that this work analyzes the performance in terms of average BLER, which is more suitable for SPC than widely-used performance metrics such as ergodic capacity and outage capacity \cite{Durisi2016,Pol2010}.
	\item Thirdly, we derive asymptotic expressions for the average BLER in the high SNR regime and carry out an analysis of diversity order, minimum blocklength and optimal power allocation for SPC-based MIMO NOMA system based on the asymptotic average BLER.
	\item Finally, we perform the blocklength comparison between MIMO NOMA and MIMO OMA systems to clarify the superiority of MIMO NOMA compared to MIMO OMA in terms of low-latency transmission when considering SPC.
\end{itemize}

\subsection{Paper Structure and Notations}

The remainder of the paper is organized as follows. Section II depicts in detail the system model and the proposed schemes. Section III presents the performance analysis in terms of average BLER for the investigated scenarios. Section IV describes an analysis of the asymptotic average BLER, diversity order, optimal power allocation, and minimum blocklength. Section V presents the numerical results. Finally, Section VI concludes this paper. For clarity, we provide a summary of main notations and symbols used in this paper in Table \ref{Table_Notations_Symbols}.

\begin{table}[!t]
	\caption{Main Notations and Symbols}
	\label{Table_Notations_Symbols}
	\centering
	\begin{tabular}{| c | m{5.9cm} |}
		\hline 
		Notation & Description \\
		\hline 
		$\left| \cdot \right|$ and $\left\| \cdot \right\|$  & The absolute value and the Euclidean norm \\
		\hline 
		${\mathcal{CN}}\left(0, N_0 \right)$  & A scalar complex Gaussian distribution with zero mean and variance $N_0$\\
		\hline 
		$Q\left(x\right)$ & The Gaussian Q-function\\
		\hline
		$\rm{Ei}\left(x\right)$ & The exponential integral function\\
		\hline
		$\Gamma\left(x,t\right)$ & The lower incomplete Gamma function \\
		\hline
		$\gamma_0$ & Average transmit signal-to-noise ratio (SNR) \\
		\hline
		$K_S$ and $K_A$ & Number of antennas at base station and user $A$ \\
		\hline
		$n_A$ & Number of information bits for user $A$ \\
		\hline
		$N_A$ & Blocklength for user $A$ \\
		\hline
		$\alpha_A$ & Power allocation coefficient for user $A$\\
		\hline
		$D_A$ & Diversity order at user $A$\\
		\hline
		$\bar \varepsilon_A$ & Average BLER at user $A$ \\
		\hline 
	\end{tabular}
\end{table}

\section{System Model \label{sys_model}}

\begin{figure}[!t]
	\centering
	\includegraphics[scale = 0.35]{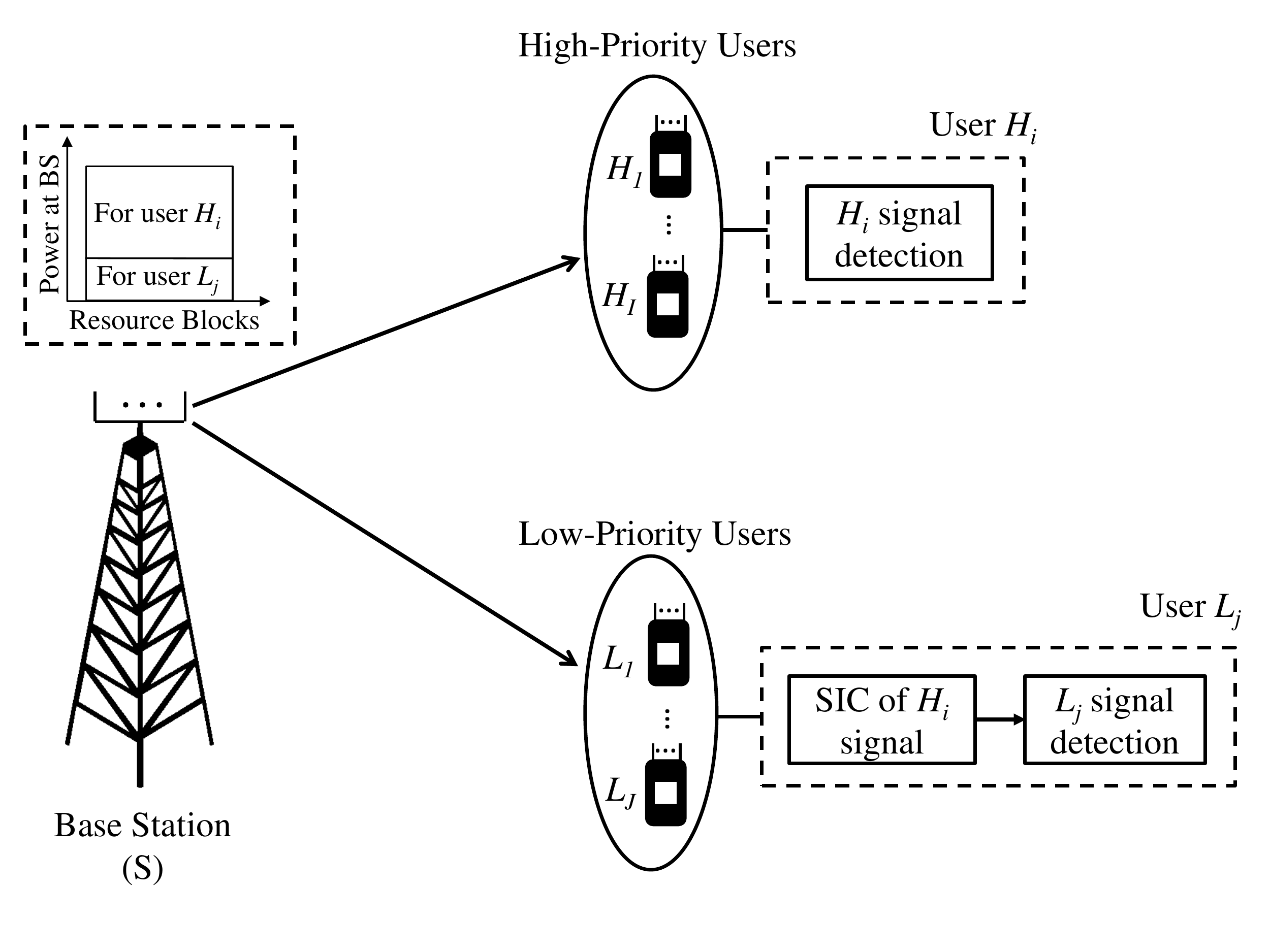}
	\caption{Model of a MIMO NOMA system under SPC over Nakagami-\textit{m} fading.} 
	\label{fig_model}
\end{figure}

In this paper, the SPC in a multiuser downlink MIMO NOMA system over Nakagami-\textit{m} fading channels is considered, as depicted in Fig. \ref{fig_model}. The network consists of one base station (BS), denoted by $S$, two cluster of users, denoted by $H = \{H_1, \dots, H_I\}$ and $L = \{L_1, \dots, L_J\}$. In addition, the BS and the users in both clusters $H$ and $L$ are equipped with $K_S$, $K_H$, and $K_L$ antennas, respectively. As reported earlier in Section \ref{sec_Intro}, it is assumed that the users' QoS requirements are taken into account in the design of the MIMO NOMA transmission in SPC instead of their channel conditions. More precisely, we consider the scenario where the users in clusters $H$ and $L$ are treated as high-priority and low-priority ones, respectively. Furthermore, the users are paired to perform NOMA with the purpose of decreasing the strong co-channel interference in NOMA systems \cite{Din2016,Liu2016,JWang2020}. Specifically, each user pair consists of two users having different priorities selected from both the clusters $H$ and $L$. Moreover, as mentioned earlier in Section \ref{sec_Intro}, to exploit the benefits of multiple antennas, we consider the scenario where TAS is employed at the BS $S$ whereas SC or MRC is utilized at the users' side (i.e., TAS/SC or TAS/MRC).

\subsection{Antenna and User Selection}
\label{subsect_tas_user_selection}

In this subsection, we present the proposed solutions of selecting antennas and users. As stated earlier, the user pairing is utilized for designing the MIMO NOMA. Specifically, the best user in each cluster is selected to perform NOMA based on the channel gains of the link from BS $S$ to the users. Furthermore, we investigate two different antenna-user selection methods, i.e., HCS and LCS, which aim to improve the performance for the users selected from clusters $H$ and $L$, respectively. It is noted that this selection process can be carried out prior to information transmission through a suitable signaling and channel state information (CSI) estimation method \cite{Yu20181,Do2018}. In addition, as in \cite{Chang2018NOMA,Yu2017,Yu20181}, the required CSI for each method is assumed to be available.

\subsubsection{HCS Method}
Since the users in cluster $H$ has higher priority than those in cluster $L$, this method focuses on improving the performance of the selected user in cluster $H$. In particular, HCS method aims to jointly select a transmit antenna and a user in cluster $H$ to maximize the channel power gain of the link from the BS $S$ to the selected user.

For the TAS/SC scheme, the indices of selected transmit antenna, $\hat k$, user and receive antenna selected from cluster $H$, $\hat i$ and $\hat{r}_H$, are given by \cite{Yan2013,Tran2018}
\begin{equation}
{\left( {\hat k,\hat i,{{\hat r}_H}} \right) = \arg \mathop {\max }\limits_{1 \le k \le {K_S},1 \le i \le I,1 \le r \le {K_H}} \left\{ {{{\left| {{h_{{S_k}{H_{i,r}}}}} \right|}^2}} \right\},}
\label{eq_kirh_tas_sc_I_def}
\end{equation}
and the indices of user and receive antenna selected from cluster $L$, $\hat j$ and $\hat{r}_L$, are expressed as
\begin{equation}
{\left( {\hat j,{{\hat r}_L}} \right) = \arg \mathop {\max }\limits_{1 \le j \le J,1 \le r \le {K_L}} \left\{ {{{\left| {{h_{{S_{\hat k}}{L_{j,r}}}}} \right|}^2}} \right\}},
\label{eq_jrl_tas_sc_I_def}
\end{equation}
where $h_{{S_k}{U_{m,n}}}$ ($U \in \left\{H, L\right\}$) denotes the channel coefficient of the link from antenna $k$ at BS $S$ to antenna $n$ at user $U_m$.

For TAS/MRC, $\hat k$, $\hat i$, and $\hat j$ are given by \cite{Yan2013}
\begin{equation}
\left( {\hat k,\hat i} \right) = \arg \mathop {\max }\limits_{1 \le k \le {N_S},1 \le i \le I} \left\{ {{{\left\| {{{\bf{h}}_{{S_k}{H_i}}}} \right\|}^2}} \right\},
\label{eq_ki_tas_mrc_I_def}
\end{equation}
and
\begin{equation}
\hat j = \arg \mathop {\max }\limits_{1 \le j \le J} \left\{ {{{\left\| {{{\bf{h}}_{{S_{\hat k}}{L_j}}}} \right\|}^2}} \right\},
\label{eq_j_tas_mrc_I_def}
\end{equation}
where ${\bf{h}}_{{S_k}{U_m}}$ represents the $N_U \times 1$ channel vector of $S_k \to U_m$ link.

\subsubsection{LCS Method}
To improve the performance of the selected user in cluster $L$ which has a lower priority, an antenna at BS $S$ and a user in cluster $L$ are jointly chosen for transmission to provide the best channel power gain of the link from BS $S$ to the selected user. Mathematically, $\hat k$, $\hat i$, $\hat j$, $\hat{r}_H$, and $\hat{r}_L$ in this method can be expressed as follows:
\begin{equation}
\begin{split}
&\text{For TAS/SC:} \\
&\left\{ {\begin{array}{*{20}{c}}
	{\left( {\hat k,\hat j,{{\hat r}_L}} \right) = \arg \mathop {\max }\limits_{1 \le k \le {K_S},1 \le j \le J,1 \le r \le {K_L}} \left\{ {{{\left| {{h_{{S_k}{L_{j,r}}}}} \right|}^2}} \right\},}\\
	{\left( {\hat i,{{\hat r}_H}} \right) = \arg \mathop {\max }\limits_{1 \le i \le I,1 \le r \le {K_H}} \left\{ {{{\left| {{h_{{S_{\hat k}}{H_{i,r}}}}} \right|}^2}} \right\},}
	\end{array}} \right.
\end{split}
\label{eq_kijrhrl_tas_sc_II_def}
\end{equation}
and for TAS/MRC:
\begin{equation}
\left\{ {\begin{array}{*{20}{c}}
	{\left( {\hat k,\hat j} \right) = \arg \mathop {\max }\limits_{1 \le k \le {N_S},1 \le j \le J} \left\{ {{{\left\| {{{\bf{h}}_{{S_k}{L_j}}}} \right\|}^2}} \right\},}\\
	{\hat i = \arg \mathop {\max }\limits_{1 \le i \le I} \left\{ {{{\left\| {{{\bf{h}}_{{S_{\hat k}}{H_i}}}} \right\|}^2}} \right\},}
	\end{array}} \right..
\label{eq_kij_tas_mrc_II_def}
\end{equation}

\subsection{Information Transmission Process and Channel Statistics}

With the NOMA protocol, BS $S$ transmits the mixed message \cite{Din2016}
\begin{equation}
x = \sqrt{P_S \alpha_{H_{\hat i}}}x_{H_{\hat i}} + \sqrt{P_S \alpha_{L_{\hat j}}}x_{L_{\hat j}}
\end{equation}
to $H_{\hat i}$ and $L_{\hat j}$. Herein, $P_S$ is the total transmit power, $\alpha_{H_{\hat i}}$ and $\alpha_{L_{\hat j}}$ ($\alpha_{H_{\hat i}} + \alpha_{L_{\hat j}} = 1$) denote the power allocation coefficients, as well as $x_{H_{\hat i}}$ and $x_{L_{\hat j}}$ represent the messages for users $H_{\hat i}$ and $L_{\hat j}$, respectively. It is noted that $\alpha_{H_{\hat i}} > \alpha_{L_{\hat j}} > 0$ due to higher priority of user $H_{\hat i}$. Thus, the received signal at user $U$ $\left(U \in \left\{H_{\hat i}, L_{\hat j}\right\}\right)$ is given by
\begin{equation}
y_U = {\bf{u}}_U{{\bf{h}}_{{S_{\hat k}}U}}\sqrt {{P_S}} \left( {\sqrt {{\alpha _{{H_{\hat i}}}}} {x_{{H_{\hat i}}}} + \sqrt {{\alpha _{{L_{\hat j}}}}} {x_{{L_{\hat j}}}}} \right) + {\bf{u}}_U{{\bf{w}}_U},
\label{yU}
\end{equation}
where ${\bf{w}}_U \sim {\cal{CN}}(0,N_0)$ denotes the additive white Gaussian noise (AWGN) at user $U$, and ${\bf{u}}_U$ represents the signal processing operation at user $U$, which is defined as in \cite{Gonzalez2016}
\begin{equation}
{{\bf{u}}_U} = \left\{ {\begin{array}{*{20}{c}}
	{{{\bf{e}}_{{K_U},{{\hat r}_U}}},}&{{\text{for TAS/SC}}}\\
	{\frac{{{\bf{h}}_{{S_{\hat k}}U}^\dag }}{{\left\| {{{\bf{h}}_{{S_{\hat k}}U}}} \right\|}},}&{{\text{for TAS/MRC}}}
	\end{array}} \right.,
\end{equation}
where ${\bf{e}}_{K,i}$ is a $1 \times K$ vector whose the $i$-th element is equal to 1, and the others are zeros.

According to NOMA principle, the user $H_{\hat i}$ can directly decode its own message, $x_{H_{\hat i}}$, since it is allocated with larger transmit power (i.e., $\alpha_{H_{\hat i}} > \alpha_{L_{\hat j}}$), hence, the interference generated by the signal of the user $L_{\hat j}$, $x_{L_{\hat j}}$, can be treated as noise \cite{Do2018}. Thus, the instantaneous signal-to-interference-plus-noise ratio (SINR) at user $H_{\hat i}$ to detect $x_{H_{\hat i}}$ is written as
\begin{equation}
\gamma_{{H_{\hat i}}}^{{x_{H_{\hat i}}}} = \frac{{{\alpha_{H_{\hat i}}}{\gamma _0}{g_{SH}}}}{{{\alpha_{L_{\hat j}}}{\gamma _0}{g_{SH}} + 1}},
\label{gHi_xh}
\end{equation}
where $\gamma_0 = \frac{P_S}{N_0}$ denotes the average transmit SNR and $g_{SH}$ is defined as
\begin{equation}
{g_{SH}} = \left\{ {\begin{array}{*{20}{c}}
	{{{\left| {{h_{{S_{\hat k}}{H_{\hat{i},\hat{r}_H}}}}} \right|}^2},}&{\text{for TAS/SC}}\\
	{{{\left\| {{{\bf{h}}_{S_{\hat k} H_{\hat i}}}} \right\|}^2},}&{\text{for TAS/MRC}}
	\end{array}} \right..
\end{equation}

Meanwhile, the user $L_{\hat j}$ utilizes SIC to first decode $x_{H_{\hat i}}$ and then remove this component from the received signal before detecting its own message, $x_{L_{\hat j}}$, \cite{Do2018}. Thus, the instantaneous SINR and SNR at $L_{\hat j}$ to detect $x_{H_{\hat i}}$ and $x_{L_{\hat j}}$ are respectively expressed as
\begin{equation}
\gamma_{{L_{\hat j}}}^{{x_{H_{\hat i}}}} = \frac{{{\alpha_{H_{\hat i}}}{\gamma _0}{g_{SL}}}}{{{\alpha_{L_{\hat j}}}{\gamma _0}{g_{SL}} + 1}},
\label{gLj_xh}
\end{equation}
and
\begin{equation}
\gamma_{{L_{\hat j}}}^{{x_{L_{\hat j}}}} = {\alpha_{L_{\hat j}}}{\gamma _0}{g_{SL}},
\label{gLj_xl}
\end{equation}
where $g_{SL}$ is given by
\begin{equation}
{g_{SL}} = \left\{ {\begin{array}{*{20}{c}}
	{{{\left| {{h_{{S_{\hat k}}{L_{\hat j,{{\hat r}_L}}}}}} \right|}^2},}&{{\text{for TAS/SC}}}\\
	{{{\left\| {{{\bf{h}}_{{S_{\hat k}}{L_{\hat j}}}}} \right\|}^2},}&{{\text{for TAS/MRC}}}
	\end{array}} \right..
\end{equation}

\section{Proposed Approach for BLER Performance Analysis with SPC}
\label{sec_per_ana}

In this section, we present some preliminaries on SPC and average BLER calculation, the derivation of CDF of channel power gains, and the average BLER analysis by utilizing HCS and LCS methods with TAS/SC and TAS/MRC schemes, specified in Section \ref{subsect_tas_user_selection}.

\subsection{Preliminaries}

Considering SPC with blocklength $N_U$ $\left(U \in \left\{H_{\hat i}, L_{\hat j}\right\}\right)$ and the number of information bits $n_U$ to user $U$, the instantaneous BLER of decoding $x_V$ $\left(V \in \left\{H_{\hat i}, L_{\hat j}\right\}\right)$ at user $U$ is approximated as in \cite{Yu2018}
\begin{equation}
\varepsilon_{U}^{x_V} \approx Q\left( {\frac{{{{\log }_2}\left( 1 + \gamma_{U}^{x_V} \right) - {n_V}/{N_V}}}{{\sqrt {v_{U}^{{x_V}}/{N_V}} }}} \right),
\label{bler_U_xV}
\end{equation}
where $Q\left( x \right) = \int\limits_x^\infty  {\frac{1}{{\sqrt {2\pi } }}{e^{ - \frac{{{t^2}}}{2}}}dt}$ denotes the Gaussian Q-function and $v_{U}^{{x_V}} = {\left( {{{\log }_2}e} \right)^2}\left[ {1 - \frac{1}{{\left( {1 + \gamma _{U}^{{x_V}}} \right)}}} \right]$ represents the channel dispersion. From (\ref{bler_U_xV}), the average BLER $\bar{\varepsilon}_{U}^{x_V}$ has the following form
\begin{equation}
\bar{\varepsilon}_{U}^{x_V} \approx \int\limits_0^\infty  {\varepsilon _U^{{x_V}}{f_{\gamma _U^{{x_V}}}}\left( x \right)dx},
\label{average_bler_U_xV_def}
\end{equation}
where $f_X (x)$ is the probability density function (PDF) of a random variable $X$. It is challenging to derive $\bar{\varepsilon}_{U}^{x_V}$ in (\ref{average_bler_U_xV_def}). Therefore, an approximation of $\varepsilon_{U}^{x_V}$ is utilized as discussed in \cite{Mak2014}, i.e.,
\begin{equation}
\varepsilon _U^{{x_V}} \approx \left\{ {\begin{array}{*{20}{c}}
	{1,}&{\gamma _U^{{x_V}} \le {v_V}}\\
	{A_U^{{x_V}},}&{{v_V} < \gamma _U^{{x_V}} < {\mu _V}}\\
	{0,}&{\gamma _U^{{x_V}} \ge {\mu _V}}
	\end{array}} \right.,
\label{bler_U_xV_approx}
\end{equation}
where $A_U^{{x_V}} = 0.5 - {\chi _{V}}\sqrt {{N_V}} \left( {\gamma _U^{{x_V}} - {\beta _{V}}} \right)$, ${\chi _{V}} = \sqrt {\frac{1}{{2\pi \left( {{2^{\frac{{2{n_V}}}{{{N_V}}}}} - 1} \right)}}}$, ${v_{V}} = {\beta _{V}} - \frac{1}{{2{\chi _{V}}\sqrt {{N_V}} }}$, ${\mu _{V}} = {\beta _{V}} + \frac{1}{{2{\chi_{V}}\sqrt {{N_V}} }}$, and ${\beta _{V}} = {2^{\frac{{{n_V}}}{{{N_V}}}}} - 1$. By substituting (\ref{bler_U_xV_approx}) into (\ref{average_bler_U_xV_def}), $\bar{\varepsilon}_{U}^{x_V}$ can be rewritten as
\begin{equation}
\bar \varepsilon _U^{{x_V}} \approx {\chi _{V}}\sqrt {{N_V}} \int\limits_{{v_{V}}}^{{\mu_{V}}} {{F_{\gamma _U^{{x_V}}}}\left( x \right)dx}.
\label{average_bler_U_xV_approx}
\end{equation}

For user $H_{\hat i}$, from (\ref{gHi_xh}) and (\ref{average_bler_U_xV_approx}), its average BLER is expressed as
\begin{equation}
\begin{split}
\bar \varepsilon _{{H_{\hat i}}} & = \bar \varepsilon _{{H_{\hat i}}}^{{x_{{H_{\hat i}}}}} \\
& \approx {\chi _{{H_{\hat i}}}}\sqrt {{N_{{H_{\hat i}}}}} \int\limits_{{v_{{H_{\hat i}}}}}^{{\mu _{{H_{\hat i}}}}} {{F_{\gamma _{{H_{\hat i}}}^{{x_{{H_{\hat i}}}}}}}\left( x \right)dx}.
\end{split}
\label{average_bler_H_def}
\end{equation}

For user $L_{\hat j}$, it needs to decode the message of user $H_{\hat i}$, $x_{H_{\hat i}}$, before detecting its own message, $x_{L_{\hat j}}$. Therefore, the average BLER at user $L_{\hat j}$ is given by
\begin{equation}
{{\bar \varepsilon }_{{L_{\hat j}}}} = \bar \varepsilon _{{L_{\hat j}}}^{{x_{{H_{\hat i}}}}} + \left( {1 - \bar \varepsilon _{{L_{\hat j}}}^{{x_{{H_{\hat i}}}}}} \right)\bar \varepsilon _{{L_{\hat j}}}^{{x_{{L_{\hat j}}}}},
\label{average_bler_L_def}
\end{equation}
where 
\begin{equation*}
\bar \varepsilon _{{L_{\hat j}}}^{{x_{{H_{\hat i}}}}} \approx {\chi _{{H_{\hat i}}}}\sqrt {{N_{{H_{\hat i}}}}} \int\limits_{{v_{{H_{\hat i}}}}}^{{\mu _{{H_{\hat i}}}}} {{F_{\gamma _{{L_{\hat j}}}^{{x_{{H_{\hat i}}}}}}}\left( x \right)dx},
\end{equation*}
and 
\begin{equation*}
\bar \varepsilon _{{L_{\hat j}}}^{{x_{{L_{\hat j}}}}} \approx {\chi _{{L_{\hat j}}}}\sqrt {{N_{{L_{\hat j}}}}} \int\limits_{{v_{{L_{\hat j}}}}}^{{\mu _{{L_{\hat j}}}}} {{F_{\gamma _{{L_{\hat j}}}^{{x_{{L_{\hat j}}}}}}}\left( x \right)dx}.
\end{equation*}

\subsection{Derivation for Cumulative Distribution Function (CDF) of Channel Power Gains}

To derive the average BLER at users $H_{\hat i}$ and $L_{\hat j}$, we first need to calculate the CDFs of $g_{SH}$ and $g_{SL}$ with TAS/SC and TAS/MRC schemes in both HCS and LCS methods. This is described as follows:
\subsubsection{HCS Method}
The CDFs of $g_{SH}$ and $g_{SL}$ with HCS method are derived in the following propositions.
\begin{proposition}
	Under HCS method and Nakagami-$m$ fading, the CDF of $g_{SH}$ with TAS/SC and TAS/MRC schemes is given by
	\begin{equation}
	{F_{{g_{SH}}}^{HCS}}\left( x \right) = 1 + \sum\limits_{p = 1}^{{a_{H,I}}} {\sum\limits_{{\Delta _H} = p} {{\Phi _H c_{H,I}}{x^{{\varphi _H}}}{e^{ - \frac{{p{m_H}x}}{{{\lambda _{SH}}}}}}} },
	\label{cdf_gSH_I}
	\end{equation}
	where $\Delta_H = \sum\nolimits_{q = 0}^{b_H - 1} {{\delta _{H,q}}}$, $\varphi_H = \sum\nolimits_{q = 0}^{b_H - 1} {q{\delta _{H,q}}}$, $\Phi_H = {\left( { - 1} \right)^p}\left[ {\prod\limits_{q = 0}^{{b_H} - 1} {{{\left( {\frac{{m_H^q}}{{q!\lambda _{SH}^q}}} \right)}^{{\delta _{H,q}}}}} } \right]$, $\lambda_{SH} = d_{SH}^{ - \theta }$, $a_{H,I} = \left\{ {\begin{array}{*{20}{c}}
		{{K_S}{K_H}I,}&{{\rm{for \enskip TAS/SC}}}\\
		{{K_S}I,}&{{\rm{for \enskip TAS/MRC}}}
		\end{array}} \right.$, $b_H = \left\{ {\begin{array}{*{20}{c}}
		{{m_H},}&{{\rm{for \enskip TAS/SC}}}\\
		{{m_H}{K_H},}&{{\rm{for \enskip TAS/MRC}}}
		\end{array}} \right.$, and $c_{H,I} = \left( {\begin{array}{*{20}{c}}
		{{a_{H,I}}}\\
		p
		\end{array}} \right)\left( {\begin{array}{*{20}{c}}
		p\\
		{{\delta _{H,0}}, \ldots ,{\delta _{H,{b_H} - 1}}}
		\end{array}} \right)$, $d_{SH}$ and $\theta$ denote the distance and path loss exponent of the link from BS $S$ to user $H_{\hat i}$, respectively.
\end{proposition}

\begin{IEEEproof}
	See Appendix A.
\end{IEEEproof}

\begin{proposition}
	Under HCS method and Nakagami-$m$ fading, the CDF of $g_{SL}$ with TAS/SC and TAS/MRC schemes is expressed as
	\begin{equation}
	{F_{{g_{SL}}}^{HCS}}\left( x \right) = 1 + \sum\limits_{p = 1}^{{a_{L,I}}} {\sum\limits_{{\Delta _L} = p} {{\Phi _L c_{L,I}}{x^{{\varphi _L}}}{e^{ - \frac{{p{m_L}x}}{{{\lambda _{SL}}}}}}} } ,
	\label{cdf_gSL_I}
	\end{equation}
	where $\Delta_L = \sum\nolimits_{q = 0}^{b_L - 1} {{\delta _{L,q}}}$, $\varphi_L = \sum\nolimits_{q = 0}^{b_L - 1} {q{\delta _{L,q}}}$, $\Phi_L = {\left( { - 1} \right)^p}\left[ {\prod\limits_{q = 0}^{{b_L} - 1} {{{\left( {\frac{{m_L^q}}{{q!\lambda _{SL}^q}}} \right)}^{{\delta _{L,q}}}}} } \right]$, $\lambda_{SL} = d_{SL}^{ - \theta }$, $a_{L,I} = \left\{ {\begin{array}{*{20}{c}}
		{{K_L}J,}&{{\rm{for \enskip TAS/SC}}}\\
		{J,}&{{\rm{for \enskip TAS/MRC}}}
		\end{array}} \right.$, $b_L = \left\{ {\begin{array}{*{20}{c}}
		{{m_L},}&{{\rm{for \enskip TAS/SC}}}\\
		{{m_L}{K_L},}&{{\rm{for \enskip TAS/MRC}}}
		\end{array}} \right.$, and $c_{L,I} = \left( {\begin{array}{*{20}{c}}
		{{a_{L,I}}}\\
		p
		\end{array}} \right)\left( {\begin{array}{*{20}{c}}
		p\\
		{{\delta _{L,0}}, \ldots ,{\delta _{L,{b_L} - 1}}}
		\end{array}} \right)$, $d_{SL}$ and $\theta$ denotes the distance of the link from BS $S$ to user $L_{\hat j}$, respectively.
\end{proposition}

\begin{IEEEproof}
	It is noted that TAS is used to select the best transmit antenna for user $H_{\hat i}$ in this case, hence, it is considered as a random solution for user $L_{\hat j}$. As such, using (\ref{eq_kirh_tas_sc_I_def}), (\ref{eq_jrl_tas_sc_I_def}), (\ref{eq_ki_tas_mrc_I_def}), and (\ref{eq_j_tas_mrc_I_def}), the CDF of $g_{SL}$ is given by \cite{Yan2013,Zha2013}
	\begin{equation}
	{F_{{g_{SL}}}}\left( x \right) = {\left( {1 - \sum\limits_{p = 0}^{{b_L} - 1} {\frac{{m_L^p}}{{p!\lambda _{SL}^p}}{x^p}{e^{ - \frac{{{m_L}x}}{{{\lambda _{SL}}}}}}} } \right)^{{a_{L,I}}}}.
	\label{cdf_gsl_I_def}
	\end{equation}
	
	By using binomial expansion and multinomial theorem similar to the proof of Proposition 1 in Appendix A, we obtain the final expression of $F_{g_{SL}} \left(x\right)$ as in (\ref{cdf_gSL_I}) and the proof is completed.
\end{IEEEproof}

\subsubsection{LCS Method}

Utilizing (\ref{eq_kijrhrl_tas_sc_II_def}), (\ref{eq_kij_tas_mrc_II_def}), and algebraic manipulations similar to the proof of Proposition 1 in Appendix A, the CDF of $g_{SH}$ and $g_{SL}$ in this case are expressed as
\begin{equation}
{F_{{g_{SH}}}^{LCS}}\left( x \right) = 1 + \sum\limits_{p = 1}^{{a_{H,II}}} {\sum\limits_{{\Delta _H} = p} {{\Phi _{H} c_{H,II}}{x^{{\varphi _H}}}{e^{ - \frac{{p{m_H}x}}{{{\lambda _{SH}}}}}}} },
\label{cdf_gSH_II}
\end{equation}
and
\begin{equation}
{F_{{g_{SL}}}^{LCS}}\left( x \right) = 1 + \sum\limits_{p = 1}^{{a_{L,II}}} {\sum\limits_{{\Delta _L} = p} {{\Phi _{L} c_{L,II}}{x^{{\varphi _L}}}{e^{ - \frac{{p{m_L}x}}{{{\lambda _{SL}}}}}}} },
\label{cdf_gSL_II}
\end{equation}
where ${a_{H,II}} = \left\{ {\begin{array}{*{20}{c}}
	{{K_H}I,}&{{\rm{for \enskip TAS/SC}}}\\
	{I,}&{{\rm{for \enskip TAS/MRC}}}
	\end{array}} \right.$, ${c_{H,II}} = \left( {\begin{array}{*{20}{c}}
	{{a_{H,II}}}\\
	p
	\end{array}} \right)\left( {\begin{array}{*{20}{c}}
	p\\
	{{\delta _{H,0}}, \ldots ,{\delta _{H,{b_H} - 1}}}
	\end{array}} \right)$, ${a_{L,II}} = \left\{ {\begin{array}{*{20}{c}}
	{{K_S}{K_L}J,}&{{\rm{for \enskip TAS/SC}}}\\
	{{K_S}J,}&{{\rm{for \enskip TAS/MRC}}}
	\end{array}} \right.$, and ${c_{L,II}} = \left( {\begin{array}{*{20}{c}}
	{{a_{L,II}}}\\
	p
	\end{array}} \right)\left( {\begin{array}{*{20}{c}}
	p\\
	{{\delta _{L,0}}, \ldots ,{\delta _{L,{b_L} - 1}}}
	\end{array}} \right)$.

\subsection{Average BLER Analysis of HCS Method}

The derivation of the average BLER at users $H_{\hat i}$ and $L_{\hat j}$ in case of using the TAS/SC or TAS/MRC scheme with HCS method are provided in the following theorems.
\begin{theorem}
	Under HCS method and Nakagami-$m$ fading, the average BLER at user $H_{\hat i}$ utilizing TAS/SC or TAS/MRC is expressed as
	\begin{equation}
	\begin{split}
	\bar \varepsilon _{{H_{\hat i}}}^{HCS} &\approx 1 + \frac{{{\chi _{{H_{\hat i}}}}{\alpha _{{H_{\hat i}}}}\sqrt {{N_{{H_{\hat i}}}}} }}{{{\gamma _0}\alpha _{{L_{\hat j}}}^2}}\sum\limits_{p = 1}^{{a_{H,I}}} {\sum\limits_{{\Delta _H} = p} {\mathop \sum \limits_{q = 0}^{{\varphi _H}} } \left( {\begin{array}{*{20}{c}}
			{{\varphi _H}}\\
			q
			\end{array}} \right)} \\
	& \times {\left( { - \frac{1}{{{\gamma _0}{\alpha _{{L_{\hat j}}}}}}} \right)^q}{\Phi _H}{c_{H,I}}{e^{\frac{{{\omega _H}}}{{{\gamma _0}{\alpha _{{L_{\hat j}}}}}}}}{\mathcal{A}_{H}},
	\end{split}
	\label{average_bler_H_I_final}
	\end{equation}
	where 
	\begin{equation*}
	{\mathcal{A}_{H}} = \left\{ {\begin{array}{*{20}{c}}
		{{\omega _H}{\Xi _{H,1}} + {\Xi _{H,2}},}&{{{\hat \varphi }_H} =  - 2}\\
		{{\rm{ - }}{\Xi _{H,1}},}&{{{\hat \varphi }_H} =  - 1}\\
		{\omega _H^{ - {{\hat \varphi }_H} - 1}{\Xi _{H,3}},}&{{{\hat \varphi }_H} \ge 0}
		\end{array}} \right.,
	\end{equation*}
	$\omega_H = \frac{pm_H}{\lambda_{SH}}$, ${\Xi_{H,1}} = {\rm{Ei}}\left( { - {\omega _H}{\phi _{{H_{\hat i}}}}} \right) - {\rm{Ei}}\left( { - {\omega _H}{\kappa _{{H_{\hat i}}}}} \right)$, ${\Xi_{H,2}} = \frac{{{e^{ - {\omega _H}{\phi _{{H_{\hat i}}}}}}}}{{{\phi _{{H_{\hat i}}}}}} - \frac{{{e^{ - {\omega _H}{\kappa _{{H_{\hat i}}}}}}}}{{{\kappa _{{H_{\hat i}}}}}}$, ${\Xi_{H,3}} = \Gamma \left( {{{\hat \varphi }_H} + 1,{\omega _H}{\phi _{{H_{\hat i}}}}} \right) - \Gamma \left( {{{\hat \varphi }_H} + 1,{\omega _H}{\kappa _{{H_{\hat i}}}}} \right)$, ${\phi _{{H_{\hat i}}}} = \frac{1}{{{\gamma _0}{\alpha _{{L_{\hat j}}}}}} + {B_{{v_{{H_{\hat i}}}}}}$, ${\kappa _{{H_{\hat i}}}} = \frac{1}{{{\gamma _0}{\alpha _{{L_{\hat j}}}}}} + {B_{{\mu _{{H_{\hat i}}}}}}$, ${B_x} = \frac{x}{{{\gamma _0}\left( {{\alpha _{{H_{\hat i}}}} - {\alpha _{{L_{\hat j}}}}x} \right)}}$, and ${{\hat \varphi }_H} = {\varphi _H} - q - 2$.
\end{theorem}

\begin{IEEEproof}
	See Appendix B.
\end{IEEEproof}

\begin{theorem}
	Under HCS method and Nakagami-$m$ fading, the average BLER at user $L_{\hat j}$ utilizing TAS/SC or TAS/MRC is given by
	\begin{equation}
	\bar \varepsilon _{{L_{\hat j}}}^{HCS} = \bar \varepsilon _{{L_{\hat j}}}^{{x_{{H_{\hat i}}},HCS}} + \left( {1 - \bar \varepsilon _{{L_{\hat j}}}^{{x_{{H_{\hat i}}}},HCS}} \right)\bar \varepsilon _{{L_{\hat j}}}^{{x_{{L_{\hat j}}}},HCS},
	\label{average_bler_L_I_final}
	\end{equation}
	where
	\begin{equation*}
	\begin{split}
	\bar \varepsilon_{{L_{\hat j}}}^{{x_{{H_{\hat i}}}},HCS} &\approx 1 + \frac{{{\chi _{{H_{\hat i}}}}{\alpha _{{H_{\hat i}}}}\sqrt {{N_{{H_{\hat i}}}}} }}{{{\gamma _0}\alpha _{{L_{\hat j}}}^2}}\sum\limits_{p = 1}^{{a_{L,I}}} {\sum\limits_{{\Delta _L} = p} {\mathop \sum \limits_{q = 0}^{{\varphi _L}} } \left( {\begin{array}{*{20}{c}}
			{{\varphi _L}}\\
			q
			\end{array}} \right)} \\
	& \times {\left( { - \frac{1}{{{\gamma _0}{\alpha _{{L_{\hat j}}}}}}} \right)^q}{\Phi _L}{c_{L,I}}{e^{\frac{{{\omega _L}}}{{{\gamma _0}{\alpha _{{L_{\hat j}}}}}}}}{{\cal A}_L},
	\end{split}
	\end{equation*}
	\begin{equation*}
	\bar \varepsilon _{{L_{\hat j}}}^{{x_{{L_{\hat j}}}},HCS} \approx 1 + {\chi _{{L_{\hat j}}}}\sqrt {{N_{{L_{\hat j}}}}} \sum\limits_{p = 1}^{{a_{L,I}}} {\sum\limits_{{\Delta _L} = p} {\frac{{{\Phi _L}{c_{L,I}}\hat \omega _L^{ - {\varphi _L} - 1}}}{{{{\left( {{\alpha _{{L_{\hat j}}}}{\gamma _0}} \right)}^{{\varphi _L}}}}}{\Xi _{L,4}}} } ,
	\end{equation*}
	\begin{equation*}
	{{\cal A}_L} = \left\{ {\begin{array}{*{20}{c}}
		{{\omega _L}{\Xi _{L,1}} + {\Xi _{L,2}},}&{{{\hat \varphi }_L} =  - 2}\\
		{{\rm{ - }}{\Xi _{L,1}},}&{{{\hat \varphi }_L} =  - 1}\\
		{\omega _L^{ - {{\hat \varphi }_L} - 1}{\Xi _{L,3}},}&{{{\hat \varphi }_L} \ge 0}
		\end{array}} \right.,
	\end{equation*}
	${\Xi _{L,1}} = {\rm{Ei}}\left( { - {\omega _L}{\phi _{{H_{\hat i}}}}} \right) - {\rm{Ei}}\left( { - {\omega _L}{\kappa _{{H_{\hat i}}}}} \right)$, ${\Xi _{L,2}} = \frac{{{e^{ - {\omega _L}{\phi _{{H_{\hat i}}}}}}}}{{{\phi _{{H_{\hat i}}}}}} - \frac{{{e^{ - {\omega _L}{\kappa _{{H_{\hat i}}}}}}}}{{{\kappa _{{H_{\hat i}}}}}}$, ${\Xi _{L,3}} = \Gamma \left( {{{\hat \varphi }_L} + 1,{\omega _L}{\phi _{{H_{\hat i}}}}} \right) - \Gamma \left( {{{\hat \varphi }_L} + 1,{\omega _L}{\kappa _{{H_{\hat i}}}}} \right)$, ${\Xi _{L,4}} = \Gamma \left( {{\varphi _L} + 1,{{\hat \omega }_L}{v_{{L_{\hat j}}}}} \right) - \Gamma \left( {{\varphi _L} + 1,{{\hat \omega }_L}{\mu _{{L_{\hat j}}}}} \right)$, ${\omega_L} = \frac{{p{m_L}}}{{{\lambda _{SL}}}}$, ${{\hat \varphi }_L} = {\varphi _L} - q - 2$, and ${{\hat \omega }_{{L}}} = \frac{{p{m_L}}}{{{\lambda _{SL}}{\alpha _{{L_{\hat j}}}}{\gamma _0}}}$.
\end{theorem}

\begin{IEEEproof}
	See Appendix C.
\end{IEEEproof}

\subsection{Average BLER Analysis of LCS Method}

In this case, the average BLER at user $H_{\hat i}$ and $L_{\hat j}$ are derived through the following theorems.
\begin{theorem}
	Under LCS method and Nakagami-$m$ fading, the average BLER at user $H_{\hat i}$ with TAS/SC or TAS/MRC is expressed as
	\begin{equation}
	\begin{split}
	\bar \varepsilon _{{H_{\hat i}}}^{LCS} &\approx 1 + \frac{{{\chi _{{H_{\hat i}}}}{\alpha _{{H_{\hat i}}}}\sqrt {{N_{{H_{\hat i}}}}} }}{{{\gamma _0}\alpha _{{L_{\hat j}}}^2}}\sum\limits_{p = 1}^{{a_{H,II}}} {\sum\limits_{{\Delta _H} = p} {\mathop \sum \limits_{q = 0}^{{\varphi _H}} } } \left( {\begin{array}{*{20}{c}}
		{{\varphi _H}}\\
		q
		\end{array}} \right)\\
	&\times {\left( { - \frac{1}{{{\gamma _0}{\alpha _{{L_{\hat j}}}}}}} \right)^q}{\Phi _H}{c_{H,II}}{e^{\frac{{{\omega _H}}}{{{\gamma _0}{\alpha _{{L_{\hat j}}}}}}}}{\mathcal{A}_H}.
	\end{split}
	\label{average_bler_H_II_final}
	\end{equation}
\end{theorem}

\begin{IEEEproof}
	To derive $\bar \varepsilon _{{H_{\hat i}}}^{LCS}$ in this theorem, the algebraic manipulations similar to the derivation of $\bar \varepsilon _{{H_{\hat i}}}^{HCS}$ in Appendix B can be utilized, where (\ref{cdf_gSH_II}) is employed instead of (\ref{cdf_gSH_I}).
\end{IEEEproof}

\begin{theorem}
	Under LCS method and Nakagami-$m$ fading, the average BLER at user $L_{\hat j}$ with TAS/SC or TAS/MRC is given by
	\begin{equation}
	{\bar \varepsilon }_{{L_{\hat j}}}^{LCS} = \bar \varepsilon _{{L_{\hat j}}}^{{x_{{H_{\hat i}}}},LCS} + \left( {1 - \bar \varepsilon _{{L_{\hat j}}}^{{x_{{H_{\hat i}}}},LCS}} \right)\bar \varepsilon _{{L_{\hat j}}}^{{x_{{L_{\hat j}}}},LCS},
	\label{average_bler_L_II_final}
	\end{equation}
	where
	\begin{equation*}
	\begin{split}
	\bar \varepsilon _{{L_{\hat j}}}^{{x_{{H_{\hat i}}}},LCS} &\approx 1 + \frac{{{\chi _{{H_{\hat i}}}}{\alpha _{{H_{\hat i}}}}\sqrt {{N_{{H_{\hat i}}}}} }}{{{\gamma _0}\alpha _{{L_{\hat j}}}^2}}\sum\limits_{p = 1}^{{a_{L,II}}} {\sum\limits_{{\Delta _L} = p} {\mathop \sum \limits_{q = 0}^{{\varphi _L}} } \left( {\begin{array}{*{20}{c}}
			{{\varphi _L}}\\
			q
			\end{array}} \right)} \\
	& \times {\left( { - \frac{1}{{{\gamma _0}{\alpha _{{L_{\hat j}}}}}}} \right)^q}{\Phi _L}{c_{L,I}}{e^{\frac{{{\omega _L}}}{{{\gamma _0}{\alpha _{{L_{\hat j}}}}}}}}{\mathcal{A}_L},
	\end{split}
	\end{equation*}
	and
	\begin{equation*}
	\bar \varepsilon _{{L_{\hat j}}}^{{x_{{L_{\hat j}}}},LCS} \approx 1 + {\chi _{{L_{\hat j}}}}\sqrt {{N_{{L_{\hat j}}}}} \sum\limits_{p = 1}^{{a_{L,II}}} {\sum\limits_{{\Delta _L} = p} {\frac{{{\Phi _L}{c_{L,II}}\hat \omega _L^{ - {\varphi _L} - 1}}}{{{{\left( {{\alpha _{{L_{\hat j}}}}{\gamma _0}} \right)}^{{\varphi _L}}}}}{\Xi _{L,4}}} }.
	\end{equation*}
\end{theorem}

\begin{IEEEproof}
	The proof of this theorem can be carried out in the same way as the proof of Theorem 2, where (\ref{cdf_gSL_II}) is used instead of (\ref{cdf_gSL_I}).
\end{IEEEproof}

\section{Proposed Analytical Framework for Optimal Power Allocation and Minimum Blocklength}

By following the average BLER analysis presented in Section \ref{sec_per_ana}, this section provides the derivation of the optimal power allocation coefficients for a minimum blocklength based on asymptotic average BLER in high SNR regime, and it also presents the analytical comparison of the minimum blocklength of NOMA with the OMA case.

\subsection{Asymptotic Average BLER Analysis}
\label{sec_asym}

As discussed in \cite{Yu2018,Zheng2019}, the average BLER, $\bar{\varepsilon}_U^{x_V}$, in (\ref{average_bler_U_xV_approx}) can be simplified by utilizing the first-order Riemann integral approximation, i.e., $\int_a^b{f(x)dx} = (b-a)f\left(\frac{a+b}{2}\right)$, as follows:
\begin{equation}
\bar{\varepsilon}_U^{x_V} \approx \chi_{V}\sqrt{N_V} \left(\mu_V - v_V\right) F_{\gamma_U^{x_V}}\left(\frac{v_V + \mu_V}{2}\right).
\label{average_bler_U_xV_Riemann_def}
\end{equation}
By substituting $v_V$ and $\mu_V$ defined in (\ref{average_bler_U_xV_def}) into (\ref{average_bler_U_xV_Riemann_def}), $\bar{\varepsilon}_U^{x_V}$ is rewritten as
\begin{equation}
\bar{\epsilon}_U^{x_V} \approx F_{\gamma_U^{x_V}}\left(\beta_V\right),
\label{average_bler_U_xV_Riemann_final}
\end{equation}
where $\beta_V$ is defined in (\ref{bler_U_xV_approx}).

By using the series representation of $e^x$ in \cite[Eq. 1.211]{Grad2007}, i.e., $e^x = \sum\limits_{k = 0}^\infty \frac{x^k}{k!}$, the asymptotic CDF of $\gamma_{H_{\hat i}}^{x_{H_{\hat i}}}$, $\gamma_{L_{\hat j}}^{x_{H_{\hat i}}}$, and $\gamma_{L_{\hat j}}^{x_{L_{\hat j}}}$ are respectively given by
\begin{equation}
F_{\gamma _{{H_{\hat i}}}^{{x_{{H_{\hat i}}}}}}^{s,\infty} \left( x \right) = F_{{g_{SH}}}^{s,\infty} \left( {{B_x}} \right)\mathop  \approx \limits^{{\gamma _0} \to \infty } \frac{{{{\left( {{m_H}{B_x}} \right)}^{{b_H}{a_{H,r}}}}}}{{{{\left( {{b_H}!} \right)}^{{a_{H,r}}}}\lambda _{SH}^{{b_H}{a_{H,r}}}}},
\label{asym_cdf_Hi_xHi_def}
\end{equation}
\begin{equation}
F_{\gamma _{{L_{\hat j}}}^{{x_{{H_{\hat i}}}}}}^{s,\infty} \left( x \right)\mathop  \approx \limits^{{\gamma _0} \to \infty } \frac{{{{\left( {{m_L}{B_x}} \right)}^{{b_L}{a_{L,r}}}}}}{{{{\left( {{b_L}!} \right)}^{{a_{L,r}}}}\lambda _{SL}^{{b_L}{a_{L,r}}}}},
\label{asym_cdf_Lj_xHi_def}
\end{equation}
and
\begin{equation}
F_{\gamma _{{L_{\hat j}}}^{{x_{{L_{\hat j}}}}}}^{s,\infty} \left( x \right)\mathop  \approx \limits^{{\gamma _0} \to \infty } \frac{{{{\left( {{m_L}{{\hat B}_x}} \right)}^{{b_L}{a_{L,r}}}}}}{{{{\left( {{b_L}!} \right)}^{{a_{L,r}}}}\lambda _{SL}^{{b_L}{a_{L,r}}}}},
\label{asym_cdf_Lj_xLj_def}
\end{equation}
where $s \in \{HCS,LCS\}$, $r = \left\{ {\begin{array}{*{20}{c}}
	{I,}&{{\rm{if }}\;s = HCS}\\
	{II,}&{{\rm{if }}\;s = LCS}
	\end{array}} \right.$, ${B_x} = \frac{x}{{{\gamma _0}\left( {{\alpha _{{H_{\hat i}}}} - {\alpha _{{L_{\hat j}}}}x} \right)}}$, and ${{\hat B}_x} = \frac{x}{{{\alpha _{{L_{\hat j}}}}{\gamma _0}}}$. From (\ref{average_bler_U_xV_Riemann_def}) - (\ref{asym_cdf_Lj_xLj_def}), the asymptotic average BLER at users $H_{\hat i}$ and $L_{\hat j}$ are respectively expressed as
\begin{equation}
\bar \varepsilon _{{H_{\hat i}}}^{s,\infty}  \approx \frac{{{{\left( {{m_H}{B_{{\beta _{{H_{\hat i}}}}}}} \right)}^{{b_H}{a_{H,r}}}}}}{{{{\left( {{b_H}!} \right)}^{{a_{H,r}}}}\lambda _{SH}^{{b_H}{a_{H,r}}}}},
\label{asym_average_bler_Hi}
\end{equation}
and
\begin{equation}
\begin{split}
\bar \varepsilon _{{L_{\hat j}}}^{s,\infty}  &= \bar \varepsilon _{{L_{\hat j}},\infty}^{{x_{{H_{\hat i}}}},s } + \left( {1 - \bar \varepsilon _{{L_{\hat j}},\infty}^{{x_{{H_{\hat i}}}},s }} \right)\bar \varepsilon _{{L_{\hat j}},\infty}^{{x_{{L_{\hat j}}}},s } \\
& \approx \bar \varepsilon _{{L_{\hat j}},\infty}^{{x_{{H_{\hat i}}}},s } + \bar \varepsilon _{{L_{\hat j}},\infty}^{{x_{{L_{\hat j}}}},s } \\
& \approx \frac{{{{\left( {{m_L}{B_{{\beta _{{H_{\hat i}}}}}}} \right)}^{{b_L}{a_{L,r}}}}}}{{{{\left( {{b_L}!} \right)}^{{a_{L,r}}}}\lambda _{SL}^{{b_L}{a_{L,r}}}}} + \frac{{{{\left( {{m_L}{{\hat B}_{{\beta _{{L_{\hat j}}}}}}} \right)}^{{b_L}{a_{L,r}}}}}}{{{{\left( {{b_L}!} \right)}^{{a_{L,r}}}}\lambda _{SL}^{{b_L}{a_{L,r}}}}}.
\end{split}
\label{asym_average_bler_Lj}
\end{equation}

From \eqref{asym_average_bler_Hi} and \eqref{asym_average_bler_Lj}, the diversity order at users $H_{\hat i}$ and $L_{\hat j}$ are respectively given by \cite{Liu2016}
\begin{equation}
\begin{split}
{D_{{H_{\hat i}}}} &=  - \mathop {\lim }\limits_{{\gamma _0} \to \infty } \frac{{\log \left( {\bar \varepsilon _{{H_{\hat i}}}^{s,\infty} } \right)}}{{\log \left( {{\gamma _0}} \right)}} \\
& = \left\{ {\begin{array}{*{20}{c}}
	{{m_H}{K_S}{K_H}I,}&{{\text{for HCS method}}}\\
	{{m_H}{K_H}I,}&{{\text{for LCS method}}}
	\end{array}} \right.,
\end{split}
\label{D_Hi}
\end{equation}
and
\begin{equation}
\begin{split}
{D_{{L_{\hat j}}}} &=  - \mathop {\lim }\limits_{{\gamma _0} \to \infty } \frac{{\log \left( {\bar \varepsilon _{{L_{\hat j}}}^{s,\infty} } \right)}}{{\log \left( {{\gamma _0}} \right)}} \\
& = \left\{ {\begin{array}{*{20}{c}}
	{{m_L}{K_L}J,}&{{\text{for HCS method}}}\\
	{{m_L}{K_S}{K_L}J,}&{{\text{for LCS method}}}
	\end{array}} \right..
\end{split}
\label{D_Lj}
\end{equation}

\begin{remark}
	For both TAS/SC and TAS/MRC schemes, the diversity orders at users $H_{\hat i}$ and $L_{\hat j}$, denoted by $\left(D_{H_{\hat i}}, D_{L_{\hat j}} \right)$, are $\left(m_H K_S K_H I, m_L K_L J \right)$ for HCS method, and $\left(m_H K_H I, m_L K_S K_L J \right)$ for LCS method. This reveals that the users $H_{\hat i}$ and $L_{\hat j}$ have achieved full diversity order with HCS and LCS methods, respectively. Furthermore, the system performance of user $H_{\hat i}$ can be improved by increasing $m_H$, $K_S$, $K_H$, and $I$ with HCS method, and by increasing $m_H$, $K_H$, and $I$ with LCS method. Meanwhile, the growth of $m_L$, $K_L$, and $J$ with HCS method, and $m_L$, $K_S$, $K_L$, and $J$ with LCS method can help enhancing the system performance of user $L_{\hat j}$.
\end{remark}

\subsection{Power and Blocklength Optimization at High SNR}
\label{sec_power_blocklength_opti_high_snr}

To determine the values of power allocation coefficients (i.e., $\alpha_{H_{\hat i}}$ and $\alpha_{L_{\hat j}}$) at which a minimum blocklength $N_U$ $\left(U \in \left\{H_{\hat i}, L_{\hat j}\right\} \right)$ is achieved to guarantee the reliability target $\bar \varepsilon_U^{th}$, the following problem needs to be addressed
\begin{mini!}[2]
{\alpha_{H_{\hat i}}, \alpha_{L_{\hat j}}}{N_U}
{\label{minN_def}}{}
\addConstraint{{{\bar \varepsilon }_U} \le \bar \varepsilon _U^{th}}{\protect \label{minN_C1}}
\addConstraint{{\alpha _{{H_{\hat i}}}} + {\alpha_{{L_{\hat j}}}} = 1, \;0 < {\alpha _{{L_{\hat j}}}} < 0.5.}{\protect \label{minN_C2}}
\end{mini!}
It is noted that $\alpha_{H_{\hat i}} = 1 - \alpha_{L_{\hat j}}$ and $\bar \varepsilon_U$ is a decreasing function of $N_U$. The problem in \eqref{minN_def} can be simplified as
\begin{mini!}[2]
	{\alpha_{L_{\hat j}}}{N_U}
	{\label{minN_simp}}{}
	\addConstraint{{{\bar \varepsilon }_U} = \bar \varepsilon _U^{th}}{\protect \label{minN_simp_C1}}
	\addConstraint{0 < {\alpha _{{L_{\hat j}}}} < 0.5.}{\protect \label{minN_simp_C2}}
\end{mini!}

By substituting \eqref{asym_average_bler_Hi} into \eqref{minN_simp_C1} for user $H_{\hat i}$ and \eqref{asym_average_bler_Lj} into \eqref{minN_simp_C1} for user $L_{\hat j}$, the blocklengths of users $H_{\hat i}$ and $L_{\hat j}$ with $s$ $\left(s \in \left\{HCS, LCS \right\}\right)$ method are respectively calculated as
\begin{equation}
{N_{{H_{\hat i}},s}} = \frac{{{n_{{H_{\hat i}}}}}}{{{{\log }_2}\left( {\frac{{1 + {{\left( {\bar \varepsilon _{{H_{\hat i}},r}^{th}/{\eta _{H,r}}} \right)}^{1/{b_H}{a_{H,r}}}}}}{{1 + {\alpha _{{L_{\hat j}}}}{{\left( {\bar \varepsilon _{{H_{\hat i}},r}^{th}/{\eta _{H,r}}} \right)}^{1/{b_H}{a_{H,r}}}}}}} \right)}},
\label{N_Hi_r}
\end{equation}
and
\begin{equation}
\begin{split}
&{N_{{L_{\hat j}},s}} \\
&= \frac{{{n_{{L_{\hat j}}}}}}{{{{\log }_2}\left( {1 + {\alpha _{{L_{\hat j}}}}{\gamma _0}{{\left( {\frac{{\bar \varepsilon _{{L_{\hat j}},r}^{th} - {\eta _{L,r}}{{\left( {\bar \varepsilon _{{H_{\hat i}},r}^{th}} \right)}^{\frac{{{b_L}{a_{L,r}}}}{{{b_H}{a_{H,r}}}}}}}}{{{{\hat \eta }_{L,r}}}}} \right)}^{1/{b_L}{a_{L,r}}}}} \right)}},
\end{split}
\label{N_Lj_r}
\end{equation}
where $\eta_{H,r} = \frac{{m_H^{{b_H}{a_{H,r}}}}}{{{{\left( {{b_H}!} \right)}^{{a_{H,r}}}}\lambda _{SH}^{{b_H}{a_{H,r}}}\gamma _0^{{b_H}{a_{H,r}}}}}$, $\eta _{L,r} = \frac{{{{\left( {{b_H}!} \right)}^{\frac{{{b_L}{a_{L,r}}}}{{{b_H}}}}}}}{{{{\left( {{b_L}!} \right)}^{{a_{L,r}}}}}}{\left( {\frac{{{m_L}{\lambda _{SH}}}}{{{m_H}{\lambda _{SL}}}}} \right)^{{b_L}{a_{L,r}}}}$, and ${{\hat \eta }_{L,r}} = \frac{{m_L^{{b_L}{a_{L,r}}}}}{{{{\left( {{b_L}!} \right)}^{{a_{L,r}}}}\lambda _{SL}^{{b_L}{a_{L,r}}}}}$.

From \eqref{N_Hi_r} and \eqref{N_Lj_r}, the derivative of $N_{{H_{\hat i}},s}$ and $N_{{L_{\hat j}},s}$ with respect to $\alpha_{L_{\hat j}}$ are derived as
\begin{equation}
\frac{{\partial {N_{{H_{\hat i}},s}}}}{{\partial {\alpha _{{L_{\hat j}}}}}} = \frac{{{n_{{H_{\hat i}}}}{\tau _{H,r}}}}{{\left( {1 + {\alpha _{{L_{\hat j}}}}{\tau _{H,r}}} \right){{\left[ {{{\log }_2}\left( {\frac{{1 + {\tau _{H,r}}}}{{1 + {\alpha _{{L_{\hat j}}}}{\tau _{H,r}}}}} \right)} \right]}^2}\ln 2}} > 0,
\label{N_Hi_r_derivative}
\end{equation}
and
\begin{equation}
\frac{{\partial {N_{{L_{\hat j}},s}}}}{{\partial {\alpha _{{L_{\hat j}}}}}} =  - \frac{{{n_{{L_{\hat j}}}}{\tau _{L,r}}}}{{\left( {1 + {\alpha _{{L_{\hat j}}}}{\tau _{L,r}}} \right){{\left[ {{{\log }_2}\left( {1 + {\alpha _{{L_{\hat j}}}}{\tau _{L,r}}} \right)} \right]}^2}\ln 2}} < 0,
\label{N_Lj_r_derivative}
\end{equation}
where ${\tau _{H,r}} = {\left( {\bar \varepsilon _{{H_{\hat i}},r}^{th}/{\eta _{H,r}}} \right)^{1/{b_H}{a_{H,r}}}}$ and $\tau_{L,r} = {\gamma _0}{\left[ {\frac{{\bar \varepsilon _{{L_{\hat j}},r}^{th} - {\eta _{L,r}}{{\left( {\bar \varepsilon _{{H_{\hat i}},r}^{th}} \right)}^{\frac{{{b_L}{a_{L,r}}}}{{{b_H}{a_{H,r}}}}}}}}{{{{\hat \eta }_{L,r}}}}} \right]^{1/{b_L}{a_{L,r}}}}$. Thus, $N_{{H_{\hat i}},s}$ is an increasing function of $\alpha_{L_{\hat j}}$, whereas $N_{{L_{\hat j}},s}$ is a decreasing function of $\alpha_{L_{\hat j}}$. Therefore, to guarantee both reliability targets $\bar \varepsilon_{{H_{\hat i}},r}^{th}$ and $\bar \varepsilon_{{L_{\hat j}},r}^{th}$, the minimum blocklength is obtained by addressing $N_{{H_{\hat i}},s} = N_{{L_{\hat j}},s} = N_{opt,s}$ and the problem of minimizing blocklength in \eqref{minN_simp} is rewritten as
\begin{mini!}[2]
	{\alpha_{L_{\hat j}}}{N_{opt,s}}
	{\label{minN_opt}}{}
	\addConstraint{{{\bar \varepsilon }_{H_{\hat i}}^s} = \bar \varepsilon_{H_{\hat i},r}^{th}}{\protect \label{minN_opt_C1}}
	\addConstraint{{{\bar \varepsilon }_{L_{\hat j},s}} = \bar \varepsilon_{L_{\hat j},r}^{th}}{\protect \label{minN_opt_C2}}
	\addConstraint{0 < {\alpha _{{L_{\hat j}}}} < 0.5.}{\protect \label{minN_opt_C3}}
\end{mini!}

Given this context, the optimal power allocation coefficient $\alpha_{L_{\hat j},opt}$ to minimize $N_{opt,s}$ can be achieved by solving the equation $f\left(\alpha_{L_{\hat j}}\right) = N_{{L_{\hat j}},s} - N_{{H_{\hat i}},s} = 0$, which is addressed in Algorithm \ref{alg_alphalj}. The minimum blocklength $N_{opt,r}$ is attained by substituting $\alpha_{L_{\hat j},opt}$ into \eqref{N_Hi_r} as follows:
\begin{equation}
{N_{opt,s}} = \frac{{{n_{{H_{\hat i}}}}}}{{{{\log }_2}\left( {\frac{{1 + {{\left( {\bar \varepsilon _{{H_{\hat i}},r}^{th}/{\eta _{H,r}}} \right)}^{1/{b_H}{a_{H,r}}}}}}{{1 + {\alpha _{{L_{\hat j}},opt}}{{\left( {\bar \varepsilon _{{H_{\hat i}},r}^{th}/{\eta _{H,r}}} \right)}^{1/{b_H}{a_{H,r}}}}}}} \right)}}.
\label{N_opt_r}
\end{equation}

\begin{algorithm}[!t]
	\SetKwInOut{Data}{Data}\SetKwInOut{Result}{Result}
	
	\Data{$n_{H_{\hat i}}$, $n_{L_{\hat j}}$, $\gamma_0$, $\bar \varepsilon _{{H_{\hat i}},r}^{th}$, $\bar \varepsilon _{{L_{\hat j}},r}^{th}$, $K_S$, $K_H$, $K_L$, $I$, $J$, $\lambda_{SH}$, $\lambda_{SL}$, and tolerance $\mu$.}
	\Result{Determine optimal power allocation coefficient $\alpha_{L_{\hat j},opt}$.}
	\BlankLine
	Initialize: $\alpha_{L_{\hat j}}^{-} \leftarrow 0$, $\alpha_{L_{\hat j}}^{+} \leftarrow 0.5$, and $\hat{\alpha}_{L_{\hat j}} \leftarrow \frac{\alpha_{L_{\hat j}}^{-} + \alpha_{L_{\hat j}}^{+}}{2}$\;
	\While{$\left| {f\left( {{{\hat \alpha }_{{L_{\hat j}}}}} \right)} \right| > \mu$}{
		\eIf{$f\left( {{{\hat \alpha }_{{L_{\hat j}}}}} \right) {f\left( \alpha_{L_{\hat j}}^{-} \right)} > 0$}{Set $\alpha_{L_{\hat j}}^{-} \leftarrow {\hat \alpha }_{{L_{\hat j}}}$\;}{
			Set $\alpha_{L_{\hat j}}^{+} \leftarrow {\hat \alpha }_{{L_{\hat j}}}$\;}
		Set $\hat{\alpha}_{L_{\hat j}} \leftarrow \frac{\alpha_{L_{\hat j}}^{-} + \alpha_{L_{\hat j}}^{+}}{2}$ and compute ${f\left( {{{\hat \alpha }_{{L_{\hat j}}}}} \right)}$ based on \eqref{N_Hi_r} and \eqref{N_Lj_r}\;
	}
	Set $\alpha_{L_{\hat j},opt} \leftarrow \hat{\alpha}_{L_{\hat j}}$\;
	\textbf{Return} $\alpha_{L_{\hat j},opt}$\;
	\caption{Proposed Power Allocation Algorithm for SPC-Based MIMO NOMA System}\label{alg_alphalj}
\end{algorithm}

\subsection{Comparison to OMA}

With OMA transmission, the minimum blocklength, $N_{OMA,s}$ $\left(r \in \{I,II\} \right)$, is the summation of the minimum blocklengths for users $H_{\hat i}$ and $L_{\hat j}$, $\hat{N}_{H_{\hat i}}$ and $\hat{N}_{L_{\hat j}}$. Similar to the derivation of blocklengths for users $H_{\hat i}$ and $L_{\hat j}$ in Section \ref{sec_power_blocklength_opti_high_snr}, $N_{OMA,s}$ in the high SNR regime is calculated as
\begin{equation}
\begin{split}
{N_{OMA,s}} &= \hat{N}_{H_{\hat i}} + \hat{N}_{L_{\hat j}} \\
&= \frac{{{n_{{H_{\hat i}}}}}}{{{{\log }_2}\left[ {1 + {{\left( {\bar \varepsilon _{{H_{\hat i}},r}^{th}/{\eta _{H,r}}} \right)}^{1/{b_H}{a_{H,r}}}}} \right]}} \\
&\quad + \frac{{{n_{{L_{\hat j}}}}}}{{{{\log }_2}\left[ {1 + {\gamma _0}{{\left( {\bar \varepsilon _{{L_{\hat j}},r}^{th}/{{\hat \eta }_{L,r}}} \right)}^{1/{b_L}{a_{L,r}}}}} \right]}}.
\end{split}
\label{N_OMA_r}
\end{equation}

From \eqref{N_opt_r} and \eqref{N_OMA_r}, the blocklength gap between NOMA and OMA, $\Delta N_s$, is given by
\begin{equation}
\Delta N_r = N_{OMA,r} - N_{opt,r} \approx \hat{N}_{H_{\hat i},r} > 0.
\label{Delta_Nr}
\end{equation}
Thus, OMA transmission needs a longer blocklength than NOMA transmission to serve the users $H_{\hat i}$ and $L_{\hat j}$.

\section{Numerical results}

In this section, we provide numerical results in terms of average BLER and minimum blocklength to characterize the effects of the proposed protocols, i.e., HCS and LCS methods with TAS/SC and TAS/MRC schemes discussed in Section \ref{subsect_tas_user_selection}, on the system performance in designing an SPC-based MIMO NOMA network. It is noted that the analysis of these performance metrics have practical significance for the reliability and latency performance evaluation of wireless systems \cite{Yu2018,Zheng2019,Lai2019,Wang2020,Xiao2019,Huang2019}. The predetermined simulation parameters are set as follows \cite{Yu2018,Zheng2019,Lai2019}: the number of information bits $n_{H_{\hat i}} = n_{L_{\hat j}} = 80$ bits; the blocklength $N_{H_{\hat i}} = N_{L_{\hat j}} = 100$; the path loss exponent $\theta = 2.5$; the distances $d_{SH} = d_{SL} = 5$ (m); the power allocation coefficients $\alpha_{H_{\hat i}} = 0.7$, and $\alpha_{H_{\hat i}} = 0.3$; the reliability targets $\bar \varepsilon_{H_{\hat i}}^{th} = 10^{-7}$ and $\bar \varepsilon_{L_{\hat j}}^{th} = 10^{-6}$.

\begin{figure}[!t]
	\centering
	\includegraphics[scale = 0.42]{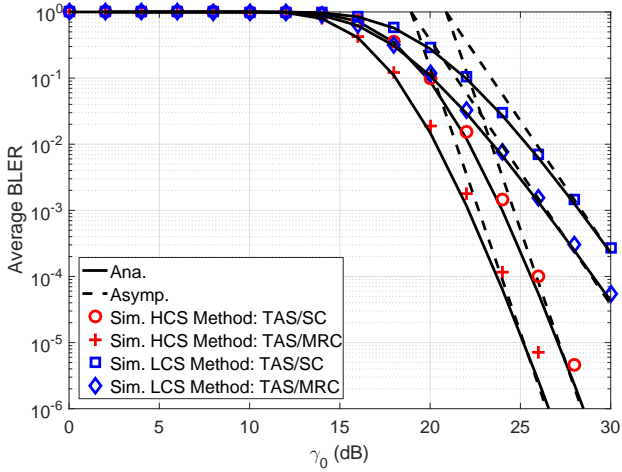}
	\caption{Average BLER at user $H_{\hat i}$ vs. $\gamma_0$ with different methods, where $m_H = m_L = 2$ and $\left( K_S, K_H, I \right) = \left(2, 2, 1\right)$.}
	\label{fig_blerH_12}
\end{figure}

\begin{figure}[!t]
	\centering
	\includegraphics[scale = 0.42]{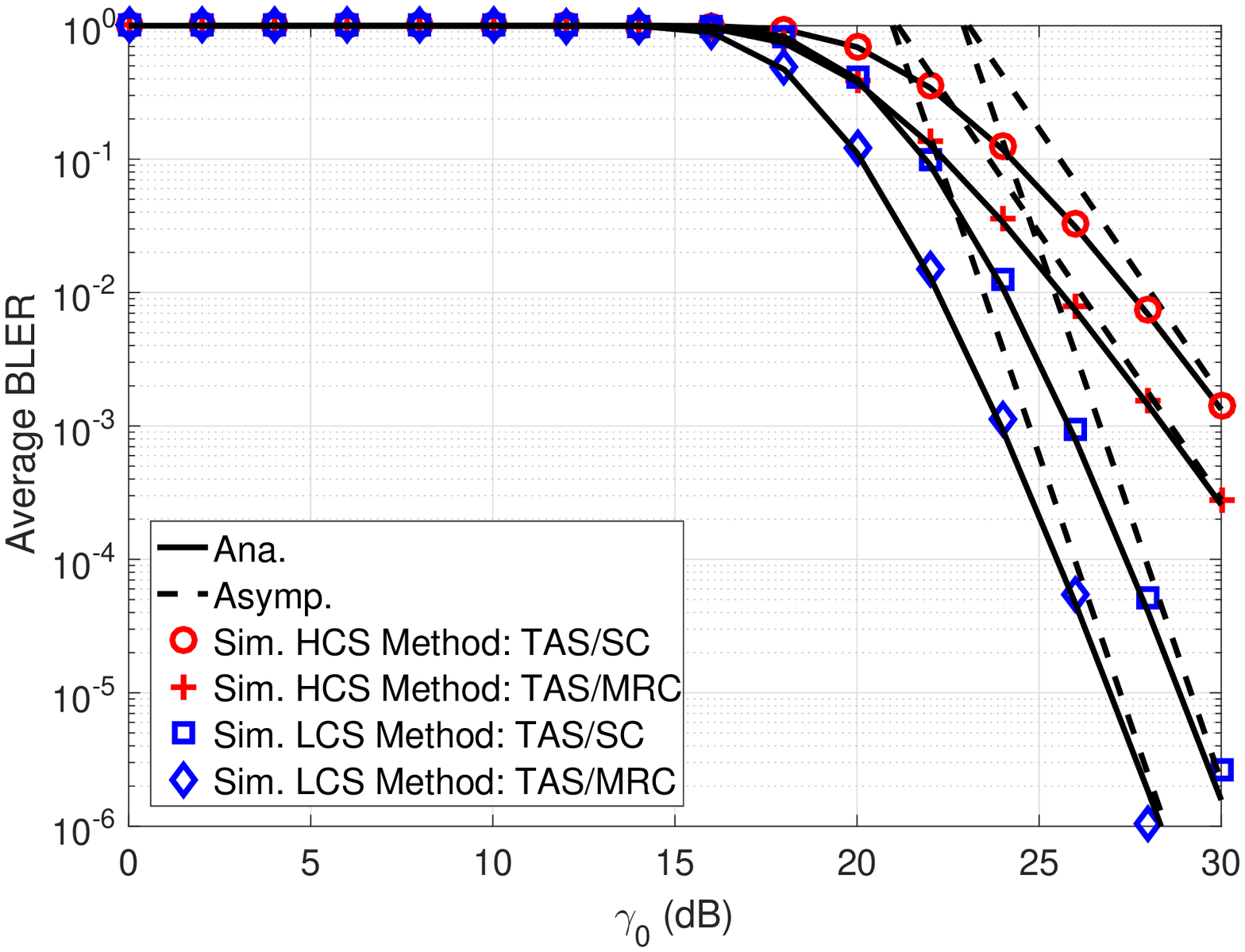}
	\caption{Average BLER at user $L_{\hat j}$ vs. $\gamma_0$ with different methods, where $m_H = m_L = 2$ and $\left( K_S, K_L, J \right) = \left(2, 2, 1\right)$.}
	\label{fig_blerL_12}
\end{figure}

In Figs. \ref{fig_blerH_12} and \ref{fig_blerL_12}, we plot the average BLERs at users $H_{\hat i}$ and $L_{\hat j}$ as a function of $\gamma_0$ with different methods (i.e., HCS method with TAS/SC or TAS/MRC and LCS method with TAS/SC or TAS/MRC). As can be observed from these figures, the analytical results are almost in good agreement with the simulation results, and the asymptotic curves accurately predict the system performance trend in the higher $\gamma_0$ regime. This verifies the correctness of our analysis in Section \ref{sec_per_ana}. In addition, Figs. \ref{fig_blerH_12} and \ref{fig_blerL_12} show that HCS method achieves better performance (i.e., lower value of average BLER is observed) for user $H_{\hat i}$ over LCS method, whereas LCS method outperforms HCS method in terms of the system performance for user $L_{\hat j}$. This result is achieved based on the fact that HCS and LCS methods are proposed to improve the received signal quality at users $H_{\hat i}$ and $L_{\hat j}$, respectively, as discussed in Section \ref{subsect_tas_user_selection}. Furthermore, these figures indicate that TAS/MRC scheme is better than TAS/SC in improving the system performance.

\begin{figure}[!t]
	\centering
	\includegraphics[scale = 0.42]{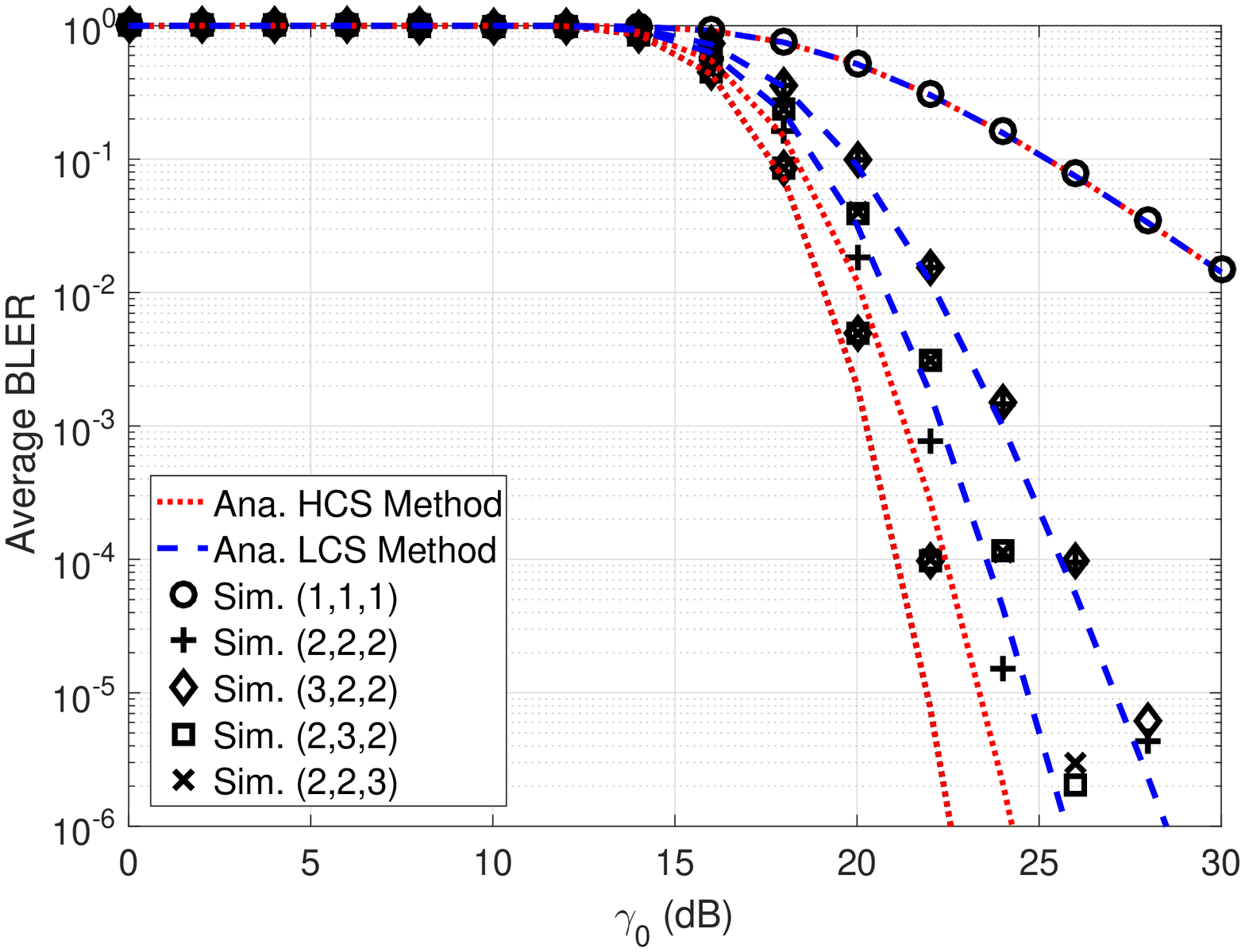}
	\caption{Average BLER at user $H_{\hat i}$ vs. $\gamma_0$ with different values of $\left( K_S, K_H, I \right)$, where $m_H = m_L = 2$.}
	\label{fig_blerH_kij}
\end{figure}

\begin{figure}[!t]
	\centering
	\includegraphics[scale = 0.42]{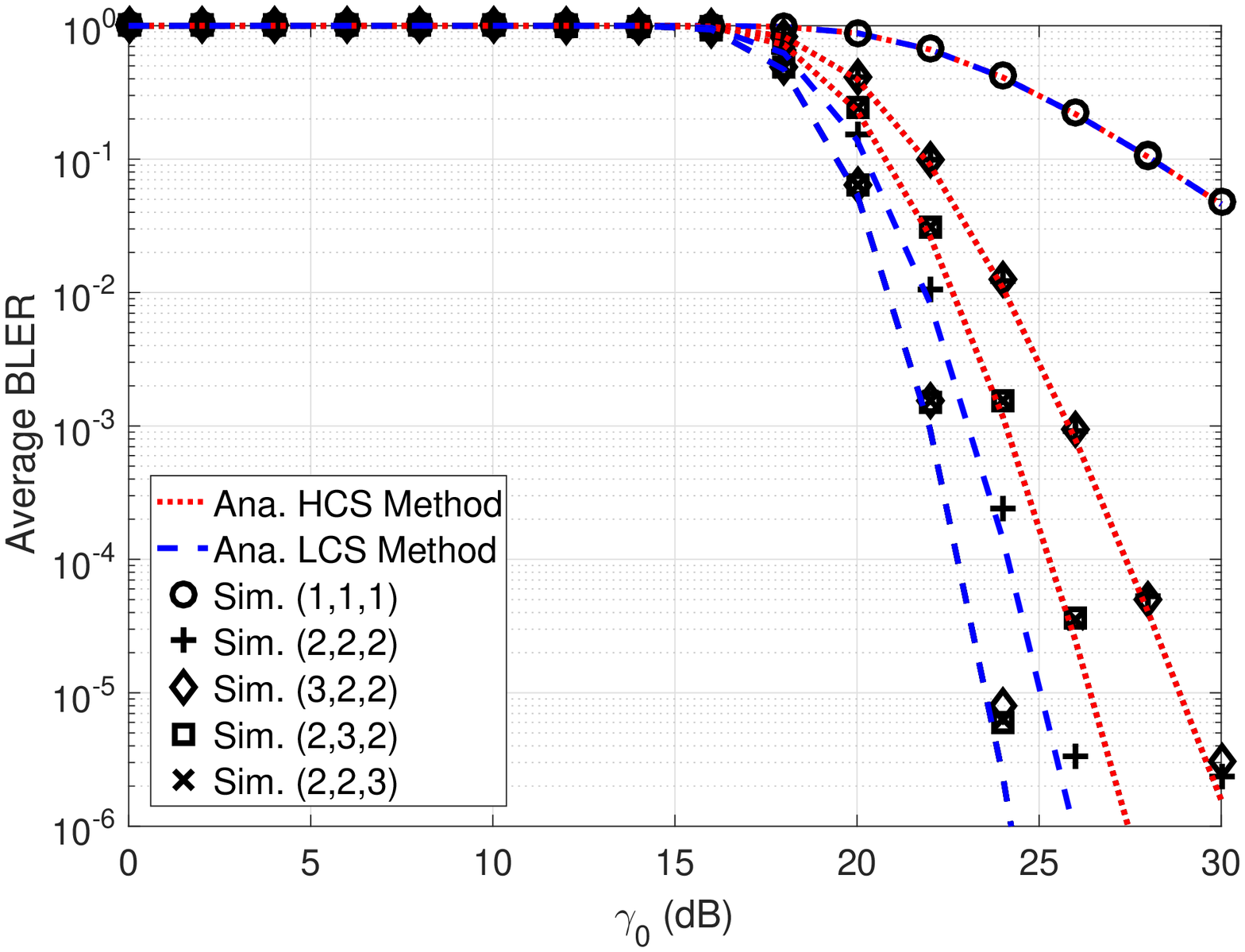}
	\caption{Average BLER at user $L_{\hat j}$ vs. $\gamma_0$ with different values of $\left( K_S, K_L, J \right)$, where $m_H = m_L = 2$.}
	\label{fig_blerL_kij}
\end{figure}

\begin{figure}[!t]
	\centering
	\includegraphics[scale = 0.42]{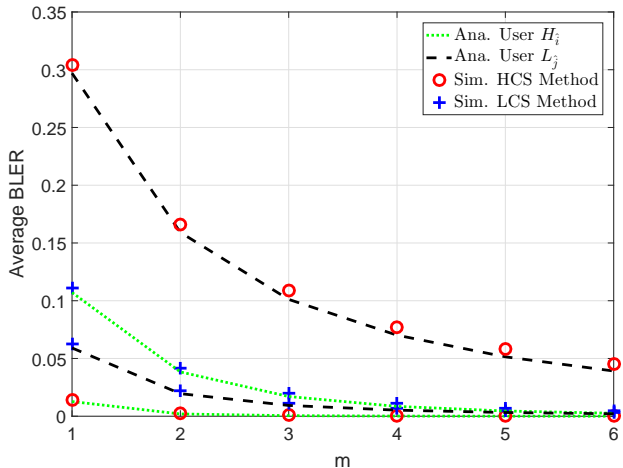}
	\caption{Average BLER at users $H_{\hat i}$ and $L_{\hat j}$ vs. $m$ with different methods, where $\gamma_0 = 20$ (dB) and $\left( K_S, K_H, K_L, I, J \right) = \left(2, 2, 2, 1, 1\right)$.}
	\label{fig_blerHL_m}
\end{figure}

\begin{figure}[!t]
	\centering
	\includegraphics[scale = 0.42]{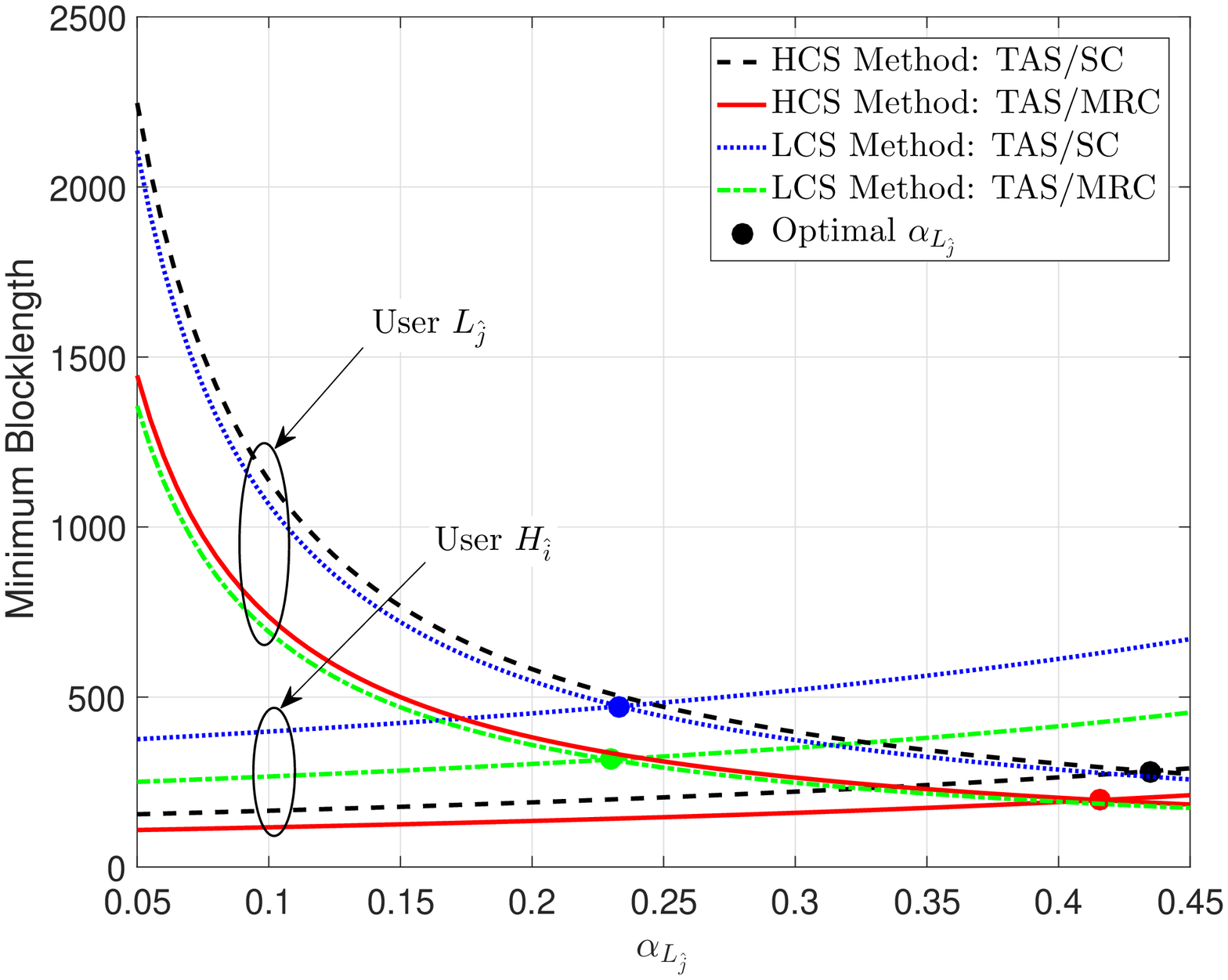}
	\caption{Minimum blocklength for users $H_{\hat i}$ and $L_{\hat j}$ vs. $\alpha_{L_{\hat j}}$ with different methods, where $m_h = m_L = 2$, $K_S = K_H = K_L = I = J = 2$, and $\gamma_0 = 20$ (dB).}
	\label{fig_Nhl_alphalj}
\end{figure}

In Figs. \ref{fig_blerH_kij} and \ref{fig_blerL_kij}, we investigate the effects of the number of users at clusters $H$ ($I$) and $L$ ($J$), and the number of antennas at BS $S$ ($K_S$), users $H_{\hat i}$ ($K_H$), and $L_{\hat j}$ ($K_L$), on the system performance. Specifically, Fig. \ref{fig_blerH_kij} shows the variation of average BLER at user $H_{\hat i}$ with respect to $\gamma_0$ with different values of $K_S$, $K_H$, and $I$, denoted by $\left( K_S, K_H, I \right)$, in case of utilizing HCS and LCS methods with the TAS/SC scheme. Meanwhile, Fig. \ref{fig_blerL_kij} plots the average BLER at user $L_{\hat j}$ versus $\gamma_0$ with different values of $\left( K_S, K_L, J \right)$ when using HCS and LCS methods with the TAS/SC scheme. These two figures indicate that as $K_S$, $K_H$, $K_L$, $I$, and $J$ are all equal to one, HCS and LCS methods result in the same curves. Furthermore, the system performance can be significantly improved by increasing $\left( K_S, K_H, I \right)$ for user $H_{\hat i}$ and $\left( K_S, K_L, J \right)$ for user $L_{\hat j}$. It is noted that the variation of $K_S$ in LCS method does not impact the system performance at user $H_{\hat i}$ (see Fig. \ref{fig_blerH_kij}). The same conclusion can be derived for user $L_{\hat j}$ when observing the change of $K_S$ in HCS method (see Fig. \ref{fig_blerL_kij}). The reason for this is based on the nature of HCS and LCS methods as mentioned in Section \ref{subsect_tas_user_selection} and the discussion part of Figs. \ref{fig_blerH_12} and \ref{fig_blerL_12}. This phenomenon also confirms our analysis of diversity order for users $H_{\hat i}$ and $L_{\hat j}$, as shown in Section \ref{sec_asym}.

In Fig. \ref{fig_blerHL_m}, we consider the change of average BLER at users $H_{\hat i}$ and $L_{\hat j}$ with respect to the fading parameters, i.e., $m_H$ and $m_L$, in case of using HCS and LCS methods with the TAS/SC scheme. Herein, we set $m_H = m_L = m$. We can see from this figure that the system performance can be improved with the increase in $m$ due to the better channel quality. More precisely, when $m=1$, Nakagami-\textit{m} fading corresponds to Rayleigh fading and the worst performance can be observed. In case of $m=(K+1)^{2}/(2K+1)$, it approximates the Rician fading with parameter $K$.

\begin{figure}[!t]
	\centering
	\includegraphics[scale = 0.42]{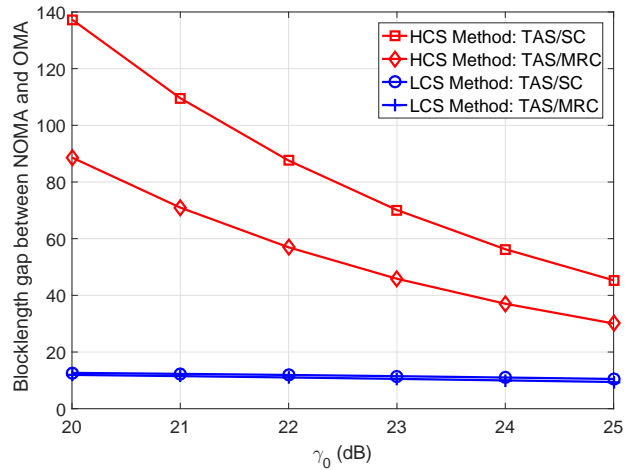}
	\caption{Blocklength comparison between NOMA and OMA.}
	\label{fig_Nhl_compare}
\end{figure}

Fig. \ref{fig_Nhl_alphalj} depicts the effect of power allocation coefficient $\alpha_{L_{\hat j}}$ on the blocklength of users $H_{\hat i}$ $\left( N_{H_{\hat i}} \right)$ and $L_{\hat j}$ $\left( N_{L_{\hat j}} \right)$. One can see from this figure that $N_{H_{\hat i}}$ and $N_{L_{\hat j}}$ are increasing and decreasing functions of $\alpha_{L_{\hat j}}$, respectively. Thus, there exists an optimal value of $\alpha_{L_{\hat j}}$, at which the minimum blocklength for both users $H_{\hat i}$ and $L_{\hat j}$ is achieved. The value of optimal $\alpha_{L_{\hat j}}$ for different cases (i.e., HCS method with TAS/SC or TAS/MRC; LCS method with TAS/SC or TAS/MRC) can be found out by using Algorithm \ref{alg_alphalj} and then the minimum blocklength is calculated by using \eqref{N_opt_r}.

In Fig. \ref{fig_Nhl_compare}, we perform the minimum blocklength comparison between NOMA and OMA transmissions ($N_{opt}$ and $N_{OMA}$) to clarify the benefits of NOMA over OMA in short-packet transmissions. As can be seen from this figure, the higher blocklength gap between NOMA and OMA, i.e., $\Delta_N$ (calculated from \eqref{Delta_Nr}), is achieved in case of using HCS method and TAS/SC scheme. This implies that the benefits of MIMO NOMA versus MIMO OMA in terms of minimum blocklength are more pronounced when utilizing HCS method as compared to LCS method. Furthermore, $\Delta_N$ is positive, hence, $N_{opt}$ is always smaller than $N_{OMA}$. In other words, MIMO NOMA can lower the transmission latency of SPC systems as compared to the MIMO OMA case.

\section{Conclusions}
In this paper, we analyzed the performance of short-packet transmission in a QoS-based multiuser downlink MIMO NOMA system over a Nakagami-\textit{m} fading channel in terms of the average BLER and minimum blocklength. Specifically, we considered the user paring to perform NOMA, where users are selected from two user clusters having different priority levels. Furthermore, we investigated different MIMO schemes including TAS for BS, SC and MRC for users, and proposed two antenna-user selection methods, i.e., HCS and LCS to design effective communication protocols for the SPC-based MIMO NOMA systems. We characterized the system performance by deriving the approximate and asymptotic (in the high SNR regime) closed-form expressions of the average BLER at the users. From the asymptotic average BLER, we carried out an analysis of diversity order, minimum blocklength, and optimal power allocation. The analytical results verified by simulation results indicated that among the proposed schemes, the HCS method with TAS/MRC and the LCS method with TAS/MRC provide the best performance with full diversity gains for the users selected from the high-priority and low-priority user clusters, respectively. Moreover, it has been demonstrated that MIMO can significantly improve the performance of NOMA systems with short-packets, and MIMO NOMA outperforms MIMO OMA in ensuring low-latency transmissions.

\appendices

\section{Proof of Proposition 1}

Using (\ref{eq_kirh_tas_sc_I_def}) and (\ref{eq_ki_tas_mrc_I_def}), the CDF of $g_{SH}$ in this case is given by \cite{Yan2013}
\begin{equation}
{F_{{g_{SH}}}^{HCS}}\left( x \right) = {\left( {1 - \sum\limits_{p = 0}^{{b_H} - 1} {\frac{{m_H^p}}{{p!\lambda _{SH}^p}}{x^p}{e^{ - \frac{{{m_H}x}}{{{\lambda _{SH}}}}}}} } \right)^{{a_{H,I}}}}.
\label{cdf_gSH_def}
\end{equation}

Applying binomial expansion in \cite[Eq. (1.111)]{Grad2007}, (\ref{cdf_gSH_def}) can be rewritten as
\begin{equation}
{F_{{g_{SH}}}^{HCS}}\left( x \right) = 1 + \sum\limits_{p = 1}^{{a_{H,I}}} \phi  \underbrace {{{\left( {\sum\limits_{q = 0}^{{b_H} - 1} {\frac{{m_H^q{x^q}}}{{q!\lambda _{SH}^q}}} } \right)}^p}}_\Psi,
\label{cdf_gSH_I_cal}
\end{equation}
where $\phi  = \left( {\begin{array}{*{20}{c}}
	{{a_{H,I}}}\\
	p
	\end{array}} \right){\left( { - 1} \right)^p}{e^{ - \frac{{p{m_H}x}}{{{\lambda _{SH}}}}}}$.

To derive (\ref{cdf_gSH_I_cal}), we first resolve $\Psi$ in (\ref{cdf_gSH_I_cal}) by utilizing the multinomial theorem as follows:
\begin{equation}
\Psi  = \sum\limits_{{\Delta _H} = p} {\psi \left[ {\prod\limits_{q = 0}^{{b_H} - 1} {{{\left( {\frac{{m_H^q}}{{q!\lambda _{SH}^q}}} \right)}^{{\delta _{H,q}}}}} } \right]{x^{{\varphi _H}}}},
\label{Psi}
\end{equation}
where $\psi  = \left( {\begin{array}{*{20}{c}}
	p\\
	{{\delta _{H,0}}, \ldots ,{\delta _{H,{b_H} - 1}}}
	\end{array}} \right)$.

The final expression of $F_{{g_{SH}}}^{HCS}\left(x\right)$ is achieved as in (\ref{cdf_gSH_I}) by substituting (\ref{Psi}) into (\ref{cdf_gSH_I_cal}).

\section{Proof of Theorem 1}

From (\ref{gHi_xh}), the CDF of $\gamma_{H_{\hat i}}^{x_{H_{\hat i}}}$ is given by
\begin{equation}
\begin{split}
{F_{\gamma _{{H_{\hat i}}}^{{x_{{H_{\hat i}}}}}}}\left( x \right) &= \Pr \left\{ {\frac{{{\alpha _{{H_{\hat i}}}}{\gamma _0}{g_{SH}}}}{{{\alpha _{{L_{\hat j}}}}{\gamma _0}{g_{SH}} + 1}} < x} \right\} \\
&= {F_{{g_{SH}}}}\left( {{B_x}} \right),
\end{split}
\label{cdf_gammash_xh}
\end{equation}
where (\ref{cdf_gammash_xh}) is obtained under the condition ${\alpha _{{H_{\hat i}}}} - {\alpha _{{L_{\hat j}}}}x > 0$ and ${B_x} = \frac{x}{{{\gamma _0}\left( {{\alpha _{{H_{\hat i}}}} - {\alpha _{{L_{\hat j}}}}x} \right)}}$ as defined in (\ref{average_bler_H_I_final}).

By substituting (\ref{cdf_gammash_xh} into (\ref{average_bler_H_def}) and using (\ref{cdf_gSH_I}), the average BLER at user $H_{\hat i}$ in HCS method with TAS/SC or TAS/MRC is expressed as
\begin{equation}
\begin{split}
\bar \varepsilon _{{H_{\hat i}}}^{HCS} &\approx 1 + {\chi _{{H_{\hat i}}}}\sqrt {{N_{{H_{\hat i}}}}} \sum\limits_{p = 1}^{{a_{H,I}}} {\sum\limits_{{\Delta _H} = p} {{\Phi _H}{c_{H,I}}} } \\
& \hspace{1cm} \times \int\limits_{{v_{{H_{\hat i}}}}}^{{\mu _{{H_{\hat i}}}}} {B_x^{{\varphi _H}}{e^{ - \frac{{p{m_H}{B_x}}}{{{\lambda _{SH}}}}}}dx},
\end{split}
\label{average_bler_H_xH_I_def}
\end{equation}
To derive the integral in (\ref{average_bler_H_xH_I_def}), we carry out the change of variable by letting $t=B_x$ and (\ref{average_bler_H_xH_I_def}) can be rewritten as
\begin{equation}
\bar \varepsilon _{{H_{\hat i}}}^{HCS} \approx 1 + {{\cal A}_{H,1}}\sum\limits_{p = 1}^{{a_{H,I}}} {\sum\limits_{{\Delta _H} = p} {{\Phi _H}{c_{H,I}}\int\limits_{{B_{{v_{{H_{\hat i}}}}}}}^{{B_{{\mu _{{H_{\hat i}}}}}}} {\frac{{{t^{{\varphi _H}}}{e^{ - \frac{{p{m_H}t}}{{{\lambda _{SH}}}}}}}}{{{{\left( {\frac{1}{{{\gamma _0}{\alpha _{{L_{\hat j}}}}}} + t} \right)}^2}}}dt} } }.
\label{average_bler_H_xH_I_cal1}
\end{equation}
By letting $u = \frac{1}{{{\gamma _0}{\alpha _{{L_{\hat j}}}}}} + t$ and using binomial expansion \cite[Eq. (1.111)]{Grad2007}, (\ref{average_bler_H_xH_I_cal1}) has the following form
\begin{equation}
\bar \varepsilon _{{H_{\hat i}}}^{HCS} \approx 1 + {{\cal{A}}_{H,1}}{{\widetilde \sum }_{H,I}}{c_{H,I}}{{\cal{A}}_{H,2}}\underbrace {\int\limits_{{\phi _{{H_{\hat i}}}}}^{{\kappa _{{H_{\hat i}}}}} {{u^{{{\hat \varphi }_H}}}{e^{ - {\omega _H}u}}du} }_{{{\cal A}_{H,3}}}.
\label{average_bler_H_xH_I_cal2}
\end{equation}
We derive $\mathcal{A}_{H,3}$ in (\ref{average_bler_H_xH_I_cal2}) with the aid of \cite[Eqs. (3.351.4), (3.352.2), and (3.351.2)]{Grad2007} and the final expression of $\bar \varepsilon _{{H_{\hat i}}}^{HCS}$ is achieved as in (\ref{average_bler_H_I_final}). 

\section{Proof of Theorem 2}

From (\ref{gLj_xh}) and (\ref{gLj_xl}), the CDF of $\gamma_{L_{\hat j}}^{x_{H_{\hat i}}}$ and $\gamma_{L_{\hat j}}^{x_{L_{\hat j}}}$ are respectively given by
\begin{equation}
\begin{split}
{F_{\gamma _{{L_{\hat j}}}^{{x_{{H_{\hat i}}}}}}}\left( x \right) &= \Pr \left\{ {\frac{{{\alpha _{{H_{\hat i}}}}{\gamma _0}{g_{SL}}}}{{{\alpha _{{L_{\hat j}}}}{\gamma _0}{g_{SL}} + 1}} < x} \right\}\\
& = {F_{{g_{SL}}}}\left( {{B_x}} \right),
\end{split}
\label{cdf_gammal_xh}
\end{equation}
and
\begin{equation}
\begin{split}
{F_{\gamma _{{L_{\hat j}}}^{{x_{{L_{\hat j}}}}}}}\left( x \right) &= \Pr \left\{ {{\alpha _{{L_{\hat j}}}}{\gamma _0}{g_{SL}} < x} \right\} \\
& = {F_{{g_{SL}}}}\left( {\frac{x}{{{\alpha _{{L_{\hat j}}}}{\gamma _0}}}} \right).
\end{split}
\label{cdf_gammal_xl}
\end{equation}

To derive ${\bar \varepsilon }_{{L_{\hat j}}}^{HCS}$ in (\ref{average_bler_L_I_final}), we need to resolve $\bar \varepsilon _{{L_{\hat j}}}^{{x_{{H_{\hat i}}}},HCS}$ and $\bar \varepsilon _{{L_{\hat j}}}^{{x_{{L_{\hat j}}}},HCS}$. For $\bar \varepsilon _{{L_{\hat j}}}^{{x_{{H_{\hat i}}}},HCS}$, from (\ref{cdf_gSL_I}), (\ref{average_bler_L_def}), and (\ref{cdf_gammal_xh}), it can be expressed as
\begin{equation}
\begin{split}
\bar \varepsilon _{{L_{\hat j}}}^{{x_{{H_{\hat i}}}},HCS} &\approx 1 + {\chi _{{H_{\hat i}}}}\sqrt {{N_{{H_{\hat i}}}}} \sum\limits_{p = 1}^{{a_{L,I}}} {\sum\limits_{{\Delta _L} = p} {{\Phi _L}{c_{L,I}}} } \\
& \times \int\limits_{{v_{{H_{\hat i}}}}}^{{\mu _{{H_{\hat i}}}}} {B_x^{{\varphi _L}}{e^{ - \frac{{p{m_L}{B_x}}}{{{\lambda _{SL}}}}}}dx}.
\end{split}
\label{average_bler_L_xH_I_cal}
\end{equation}
After some algebraic manipulations similar to the proof of Theorem 1 in Appendix B, the final expression for $\bar \varepsilon _{{L_{\hat j}}}^{{x_{{H_{\hat i}}}},HCS}$ can be obtained as in (\ref{average_bler_L_I_final}).

For $\bar \varepsilon_{{L_{\hat j}}}^{{x_{{L_{\hat j}}}},HCS}$, we derive it with the aid of (\ref{average_bler_L_def}), (\ref{cdf_gSL_I}), and (\ref{cdf_gammal_xl}) as follows:
\begin{equation}
\begin{split}
\bar \varepsilon _{{L_{\hat j}}}^{{x_{{L_{\hat j}}}},HCS} &\approx 1 + {\chi _{{L_{\hat j}}}}\sqrt {{N_{{L_{\hat j}}}}} \sum\limits_{p = 1}^{{a_{L,I}}} {\sum\limits_{{\Delta _L} = p} {\frac{{{\Phi _L}{c_{L,I}}}}{{{{\left( {{\alpha _{{L_{\hat j}}}}{\gamma _0}} \right)}^{{\varphi _L}}}}}} } \\
& \times \int\limits_{{v_{{L_{\hat j}}}}}^{{\mu _{{L_{\hat j}}}}} {{x^{{\varphi _L}}}{e^{ - {{\hat \omega }_L}x}}dx}.
\end{split}
\label{average_bler_L_xL_I_cal}
\end{equation}
By using \cite[Eq. (3.351.2)]{Grad2007}, the integral in (\ref{average_bler_L_xL_I_cal}) can be represented as
\begin{equation}
\int\limits_{{v_{{L_{\hat j}}}}}^{{\mu _{{L_{\hat j}}}}} {{x^{{\varphi _L}}}{e^{ - {{\hat \omega }_L}x}}dx}  = \hat \omega _L^{ - {\varphi _L} - 1}{\Xi _{L,4}},
\label{integral_derive}
\end{equation}
where $\Xi _{L,4}$ is defined in (\ref{average_bler_L_I_final}). By substituting (\ref{integral_derive}) into (\ref{average_bler_L_xL_I_cal}), we obtain the final expression for $\bar \varepsilon_{{L_{\hat j}}}^{{x_{{L_{\hat j}}}},HCS}$ as in (\ref{average_bler_L_I_final}).



\ifCLASSOPTIONcaptionsoff
  \newpage
\fi

\bibliographystyle{IEEEtran}
\bibliography{IEEEabrv,refs}

\begin{thebibliography}{10}
\providecommand{\url}[1]{#1}
\csname url@samestyle\endcsname
\providecommand{\newblock}{\relax}
\providecommand{\bibinfo}[2]{#2}
\providecommand{\BIBentrySTDinterwordspacing}{\spaceskip=0pt\relax}
\providecommand{\BIBentryALTinterwordstretchfactor}{4}
\providecommand{\BIBentryALTinterwordspacing}{\spaceskip=\fontdimen2\font plus
\BIBentryALTinterwordstretchfactor\fontdimen3\font minus
  \fontdimen4\font\relax}
\providecommand{\BIBforeignlanguage}[2]{{%
\expandafter\ifx\csname l@#1\endcsname\relax
\typeout{** WARNING: IEEEtran.bst: No hyphenation pattern has been}%
\typeout{** loaded for the language `#1'. Using the pattern for}%
\typeout{** the default language instead.}%
\else
\language=\csname l@#1\endcsname
\fi
#2}}
\providecommand{\BIBdecl}{\relax}
\BIBdecl

\bibitem{Ji2018}
H.~Ji, S.~Park, J.~Yeo, Y.~Kim, J.~Lee, and B.~Shim, ``Ultra-reliable and
  low-latency communications in {5G} downlink: Physical layer aspects,''
  \emph{IEEE Wireless Commun.}, vol.~25, no.~3, pp. 124--130, Jun. 2018.

\bibitem{Popovski2018}
P.~Popovski, J.~J. Nielsen, C.~Stefanovic, E.~de~Carvalho, E.~Strom, K.~F.
  Trillingsgaard, A.-S. Bana, D.~M. Kim, R.~Kotaba, J.~Park, and R.~B.
  Sorensen, ``Wireless access for ultra-reliable low-latency communication:
  Principles and building blocks,'' \emph{IEEE Netw.}, vol.~32, no.~2, pp.
  16--23, Mar. 2018.

\bibitem{Sharma2020}
S.~K. Sharma and X.~Wang, ``Toward massive machine type communications in
  ultra-dense cellular {IoT} networks: Current issues and machine
  learning-assisted solutions,'' \emph{IEEE Commun. Surveys Tuts.}, vol.~22,
  no.~1, pp. 426--471, Firstquarter 2020.

\bibitem{Sutton2019}
G.~J. Sutton, J.~Zeng, R.~P. Liu, W.~Ni, D.~N. Nguyen, B.~A. Jayawickrama,
  X.~Huang, M.~Abolhasan, Z.~Zhang, E.~Dutkiewicz, and T.~Lv, ``Enabling
  technologies for ultra-reliable and low latency communications: From {PHY}
  and {MAC} layer perspectives,'' \emph{IEEE Commun. Surveys Tuts.}, vol.~21,
  no.~3, pp. 2488--2524, Feb. 2019.

\bibitem{Durisi2016}
G.~Durisi, T.~Koch, and P.~Popovski, ``Toward massive, ultrareliable, and
  low-latency wireless communication with short packets,'' \emph{Proc. IEEE},
  vol. 104, no.~9, pp. 1711--1726, Sep. 2016.

\bibitem{Pol2010}
Y.~Polyanskiy, H.~V. Poor, and S.~Verdu, ``Channel coding rate in the finite
  blocklength regime,'' \emph{IEEE Trans. Inf. Theory}, vol.~56, no.~5, pp.
  2307--2359, May 2010.

\bibitem{Yang2014}
W.~Yang, G.~Durisi, T.~Koch, and Y.~Polyanskiy, ``Quasi-static multiple-antenna
  fading channels at finite blocklength,'' \emph{IEEE Trans. Inf. Theory},
  vol.~60, no.~7, pp. 4232--4265, Jun. 2014.

\bibitem{Mousaei2017}
M.~Mousaei and B.~Smida, ``Optimizing pilot overhead for ultra-reliable
  short-packet transmission,'' in \emph{IEEE Int. Conf. Commun. (ICC)}, Paris,
  France, May 2017.

\bibitem{Dai2015}
L.~Dai, B.~Wang, Y.~Yuan, S.~Han, C.-L. I, and Z.~Wang, ``Nonorthogonal
  multiple access for 5{G}: {S}olutions, challenges, opportunities, and future
  research trends,'' \emph{{IEEE} Commun. Mag.}, vol.~53, no.~9, pp. 74--81,
  Sep. 2015.

\bibitem{Lei2019NOMA}
L.~Lei, L.~You, Y.~Yang, D.~Yuan, S.~Chatzinotas, and B.~Ottersten, ``Load
  coupling and energy optimization in multi-cell and multi-carrier {NOMA}
  networks,'' \emph{IEEE Trans. Veh. Technol.}, vol.~68, no.~11, pp.
  11\,323--11\,337, Nov. 2019.

\bibitem{Dai2018}
L.~Dai, B.~Wang, Z.~Ding, Z.~Wang, S.~Chen, and L.~Hanzo, ``A survey of
  non-orthogonal multiple access for {5G},'' \emph{IEEE Commun. Surveys Tuts.},
  vol.~20, no.~3, pp. 2294--2323, Thirdquarter 2018.

\bibitem{Cirik2019}
A.~C. Cirik, N.~M. Balasubramanya, L.~Lampe, G.~Vos, and S.~Bennett, ``Toward
  the standardization of grant-free operation and the associated {NOMA}
  strategies in {3GPP},'' \emph{IEEE Commun. Stand. Mag.}, vol.~3, no.~4, pp.
  60--66, Dec. 2019.

\bibitem{3GPP2016}
3rd Generation Partnership Project~(3GPP), ``Study on downlink multiuser
  superposition transmission {(MUST)} for {LTE},'' 3GPP, Tech. Rep. 36.859,
  Jan. 2016.

\bibitem{3GPP042020}
------, ``Evolved universal terrestrial radio access {(E-UTRA)}; {P}hysical
  channels and modulation,'' 3GPP, Tech. Rep. TS 36.211, Apr. 2020.

\bibitem{3GPP2018}
------, ``Study on non-orthogonal multiple access {(NOMA)} for {NR},'' 3GPP,
  Tech. Rep. 38.812, Dec. 2018.

\bibitem{Sun2018}
X.~Sun, S.~Yan, N.~Yang, Z.~Ding, C.~Shen, and Z.~Zhong, ``Short-packet
  downlink transmission with non-orthogonal multiple access,'' \emph{IEEE
  Trans. Wireless Commun.}, vol.~17, no.~7, pp. 4550--4564, Jul. 2018.

\bibitem{Makki2020}
B.~Makki, K.~Chitti, A.~Behravan, and M.-S. Alouini, ``A survey of {NOMA}:
  Current status and open research challenges,'' \emph{IEEE Open J. Commun.
  Soc.}, no.~1, pp. 179--189, Jan. 2020.

\bibitem{Bud2020NOMAD2D}
I.~Budhiraja, N.~Kumar, and S.~Tyagi, ``Cross-layer interference management
  scheme for {D2D} mobile users using {NOMA},'' \emph{IEEE Syst. J.}, pp.
  1--12, Jun. 2020, {Early Access}.

\bibitem{Chang2018NOMA}
Z.~Chang, L.~Lei, H.~Zhang, T.~Ristaniemi, S.~Chatzinotas, B.~Ottersten, and
  Z.~Han, ``Secure and energy-efficient resource allocation for
  multiple-antenna {NOMA} with wireless power transfer,'' \emph{IEEE Trans.
  Green Commun. Netw.}, vol.~2, no.~4, pp. 1059--1071, Dec. 2018.

\bibitem{Yu2020}
Q.~Yu, C.~Han, L.~Bai, J.~Wang, J.~Choi, and X.~Shen, ``Multiuser selection
  criteria for {MIMO-NOMA} systems with different detectors,'' \emph{IEEE
  Trans. Veh. Technol.}, vol.~69, no.~2, pp. 1777--1791, Feb. 2020.

\bibitem{Zeng2017}
M.~Zeng, A.~Yadav, O.~A. Dobre, G.~I. Tsiropoulos, and H.~V. Poor, ``Capacity
  comparison between {MIMO-NOMA} and {MIMO-OMA} with multiple users in a
  cluster,'' \emph{IEEE J. Sel. Areas Commun.}, vol.~35, no.~10, pp.
  2413--2424, Oct. 2017.

\bibitem{Yu2017}
Y.~Yu, H.~Chen, Y.~Li, Z.~Ding, and L.~Zhuo, ``Antenna selection in {MIMO}
  cognitive radio-inspired {NOMA} systems,'' \emph{IEEE Commun. Lett.},
  vol.~21, no.~12, pp. 2658--2661, Dec. 2017.

\bibitem{Yu20181}
Y.~Yu, H.~Chen, Y.~Li, Z.~Ding, L.~Song, and B.~Vucetic, ``Antenna selection
  for {MIMO} nonorthogonal multiple access systems,'' \emph{IEEE Trans. Veh.
  Technol.}, vol.~67, no.~4, pp. 3158--3171, Apr. 2018.

\bibitem{Yu2018}
Y.~Yu, H.~Chen, Y.~Li, Z.~Ding, and B.~Vucetic, ``On the performance of
  non-orthogonal multiple access in short-packet communications,'' \emph{IEEE
  Commun. Lett.}, vol.~22, no.~3, pp. 590--593, Mar. 2018.

\bibitem{Zheng2019}
J.~Zheng, Q.~Zhang, and J.~Qin, ``Average block error rate of downlink {NOMA}
  short-packet communication systems in nakagami-m fading channels,''
  \emph{IEEE Commun. Lett.}, vol.~23, no.~10, pp. 1712--1716, Oct. 2019.

\bibitem{Lai2019}
X.~Lai, Q.~Zhang, and J.~Qin, ``Cooperative {NOMA} short-packet communications
  in flat {Rayleigh} fading channels,'' \emph{IEEE Trans. Veh. Technol.},
  vol.~68, no.~6, pp. 6182--6186, Jun. 2019.

\bibitem{Wang2020}
Z.~Wang, T.~Lv, Z.~Lin, J.~Zeng, and P.~T. Mathiopoulos, ``Outage performance
  of {URLLC NOMA} systems with wireless power transfer,'' \emph{IEEE Wireless
  Commun. Lett.}, vol.~9, no.~3, pp. 380--384, Mar. 2020.

\bibitem{Xiao2019}
C.~Xiao, J.~Zeng, W.~Ni, X.~Su, R.~P. Liu, T.~Lv, and J.~Wang, ``Downlink
  {MIMO-NOMA} for ultra-reliable low-latency communications,'' \emph{IEEE J.
  Sel. Areas Commun.}, vol.~37, no.~4, pp. 780--794, Apr. 2019.

\bibitem{Huang2019}
X.~Huang and N.~Yang, ``On the block error performance of short-packet
  non-orthogonal multiple access systems,'' in \emph{IEEE Int. Conf. Commun.
  (ICC)}, Shanghai, China, May 2019.

\bibitem{Yan2013}
N.~Yang, P.~L. Yeoh, M.~Elkashlan, R.~Schober, and I.~B. Collings, ``Transmit
  antenna selection for security enhancement in {MIMO} wiretap channels,''
  \emph{IEEE Trans. Commun.}, vol.~61, no.~1, pp. 144--154, Jan. 2013.

\bibitem{Do2018}
T.~N. Do, D.~B. d.~Costa, T.~Q. Duong, and B.~An, ``Improving the performance
  of cell-edge users in {MISO-NOMA} systems using {TAS} and {SWIPT}-based
  cooperative transmissions,'' \emph{IEEEE Trans. Green Commun. Netw.}, vol.~2,
  no.~1, pp. 49--62, Mar. 2018.

\bibitem{Ding2016nomaIoT}
Z.~Ding, L.~Dai, and H.~V. Poor, ``{MIMO-NOMA} design for small packet
  transmission in the internet of things,'' \emph{IEEE Access}, vol.~4, pp.
  1393--1405, Apr. 2016.

\bibitem{Din20162}
Z.~Ding, H.~Dai, and H.~V. Poor, ``Relay selection for cooperative {NOMA},''
  \emph{IEEE Wireless Commun. Lett.}, vol.~5, no.~4, pp. 416--419, Jun. 2016.

\bibitem{Tran2018}
D.-D. Tran, D.-B. Ha, V.~N. Vo, C.~So-In, H.~Tran, T.~G. Nguyen, Z.~A. Baig,
  and S.~Sanguanpong, ``Performance analysis of {DF/AF} cooperative {MISO}
  wireless sensor networks with {NOMA} and {SWIPT} over {N}akagami-m fading,''
  \emph{IEEE Access}, vol.~6, pp. 56\,142--56\,161, Oct. 2018.

\bibitem{Din2016}
Z.~Ding, P.~Fan, and H.~V. Poor, ``Impact of user pairing on 5{G} nonorthogonal
  multiple access downlink transmissions,'' \emph{IEEE Trans. Veh. Technol.},
  vol.~65, no.~8, pp. 6010--6023, Aug. 2016.

\bibitem{Liu2016}
Y.~Liu, Z.~Ding, M.~Elkashlan, and H.~V. Poor, ``Cooperative nonorthogonal
  multiple access with simultaneous wireless information and power transfer,''
  \emph{IEEE J. Sel. Areas Commun.}, vol.~34, no.~4, pp. 938--953, Apr. 2016.

\bibitem{JWang2020}
J.~Wang, B.~Xia, K.~Xiao, and Z.~Chen, ``Performance analysis and power
  allocation strategy for downlink {NOMA} systems in large-scale cellular
  networks,'' \emph{IEEE Trans. Veh. Technol.}, vol.~69, no.~3, pp. 3459--3464,
  Mar. 2020.

\bibitem{Gonzalez2016}
D.~C. González, D.~B. da~Costa, and J.~C. S.~S. Filho, ``Distributed {TAS/MRC}
  and {TAS/SC} schemes for fixed-gain {AF} systems with multiantenna relay:
  Outage performance,'' \emph{IEEE Trans. Wireless Commun.}, vol.~15, no.~6,
  pp. 4380--4392, Jun. 2016.

\bibitem{Mak2014}
B.~Makki, T.~Svensson, and M.~Zorzi, ``Finite block-length analysis of the
  incremental redundancy {HARQ},'' \emph{IEEE Wireless Commun. Lett.}, vol.~3,
  no.~5, pp. 529--532, Oct. 2014.

\bibitem{Zha2013}
X.~Zhang, X.~Zhou, and M.~R. McKay, ``Enhancing secrecy with multi-antenna
  transmission in wireless ad hoc networks,'' \emph{IEEE Trans. Inf. Forensics
  Security}, vol.~8, no.~11, pp. 1802--1814, Nov. 2013.

\bibitem{Grad2007}
I.~Gradshteyn and I.~Ryzhik, \emph{Table of Integrals, Series, and Products},
  7th~ed.\hskip 1em plus 0.5em minus 0.4em\relax Academic Press, Mar. 2007.

\end{thebibliography}

\end{document}